\documentclass[acmsmall,screen]{acmart}

\copyrightyear{2026}
\setcopyright{none}

\setcopyright{rightsretained}
\acmDOI{10.1145/3797116}
\acmYear{2026}
\acmJournal{PACMSE}
\acmVolume{3}
\acmNumber{FSE}
\acmArticle{FSE088}
\acmMonth{7}
\received{2025-09-11}
\received[accepted]{2025-12-22}

\usepackage{xspace}
\usepackage{enumitem}
\usepackage{url}
\usepackage{array}
\usepackage{listings}
\usepackage{algorithm}
\usepackage{algorithmic}
\usepackage{color}
\usepackage{tikz}
\usepackage[utf8]{inputenc} 
\usepackage{booktabs, multirow} 
\usepackage{soul}
\usepackage{colortbl} 
\usepackage{changepage,threeparttable} 

\usepackage[english]{babel}
\addto\extrasenglish{%
}

\newcommand{\sys}{\mbox{\textsc{asmFooler}}\xspace}

\newcommand{\cc}[1]{\mbox{\fontsize{8}{9}\selectfont\texttt{#1}}}

\newcommand{\etal}{{\em et al.}\xspace}
\newcommand{\eg}{{\em e.g.,}\xspace}
\newcommand{\ie}{{\em i.e.,}\xspace}

\newcommand{\PP}[1]{\par\addvspace{2px}\noindent{\textbf{#1.}~}\ignorespaces}

\newcommand*\WC[1]{%
	\begin{tikzpicture}[baseline=(C.base)]
		\node[draw,circle,inner sep=0.2pt](C) {#1};
\end{tikzpicture}}

\lstdefinestyle{customasm}{
    belowcaptionskip=1\baselineskip,
    frame=single, 
    frameround=tttt,
    xleftmargin=\parindent,
    language=[x86masm]Assembler, 
    morekeywords={pushfq, popfq, movl, callq, nopl},
    basicstyle=\fontsize{7}{7}\ttfamily,
    commentstyle=\itshape\color{green!60!black},
    keywordstyle=\color{blue!80!black},
    identifierstyle=\color{red!80!black},
    tabsize=4,
    numbers=left,
    numbersep=8pt,
    stepnumber=1,
    numberstyle=\tiny\color{gray}, 
    columns = fullflexible,
    escapechar=@,
    literate={806d}{{\textcolor{black}{806d}}}{1}
    {807d}{{\textcolor{black}{807d}}}{1}
    {830c}{{\textcolor{black}{830c}}}{1}
    {830d}{{\textcolor{black}{830d}}}{1}
    {830f}{{\textcolor{black}{830f}}}{1}
    {831f}{{\textcolor{black}{831f}}}{1}
    {832d}{{\textcolor{black}{832d}}}{1}
    {...}{{\textcolor{black}{...}}}{1}
    {initialize\_eval}{{\textcolor{black}{initialize\_eval}}}{1}
    {\%}{{\textcolor{red!80!black}{\%}}}{1}
    {eax}{{\textcolor{red!80!black}{eax}}}{1}
    {ecx}{{\textcolor{red!80!black}{ecx}}}{1}
    {edi}{{\textcolor{red!80!black}{edi}}}{1}
    {0xf21}{{\textcolor{red!80!black}{0xf21}}}{1}
    {0xf26}{{\textcolor{red!80!black}{0xf26}}}{1}
    {0x29751}{{\textcolor{red!80!black}{0x29751}}}{1}
    {0x5f23a}{{\textcolor{red!80!black}{0x5f23a}}}{1}
    {0x4ae23}{{\textcolor{red!80!black}{0x4ae23}}}{1}
    {0x830c}{{\textcolor{red!80!black}{0x830c}}}{1}
    {0x7xfe0}{{\textcolor{red!80!black}{0x7xfe0}}}{1}
    {0x7xfe5}{{\textcolor{red!80!black}{0x7xfe5}}}{1}
    {0x8090}{{\textcolor{red!80!black}{0x8090}}}{1}
    {8059:    movl   \$0x0,0xf21c9(\%rip)}{{\textbf{8059:    movl   \$0x0,0xf21c9(\%rip)}}}{1}
    ,
}
\lstdefinestyle{customasm2}{
    belowcaptionskip=1\baselineskip,
    frame=single, 
    frameround=tttt,
    xleftmargin=\parindent,
    language=[x86masm]Assembler,
    morekeywords={pushfq, popfq, nopl},
    basicstyle=\fontsize{7}{7}\ttfamily,
    commentstyle=\itshape\color{green!60!black},
    keywordstyle=\color{blue!80!black},
    identifierstyle=\color{red!80!black},
    tabsize=4,
    numbers=left,
    numbersep=8pt,
    stepnumber=1,
    numberstyle=\tiny\color{gray}, 
    columns = fullflexible,
    firstnumber=13,
    literate={\%}{{\textcolor{red!80!black}{\%}}}{1}
    {...}{{\textcolor{black}{...}}}{1}
    {831f}{{\textcolor{black}{831f}}}{1}
    {832d}{{\textcolor{black}{832d}}}{1}
    {\%eax}{{\textcolor{red!80!black}{\%eax}}}{1}
    {\%edi}{{\textcolor{red!80!black}{\%edi}}}{1}
    {0x29751}{{\textcolor{red!80!black}{0x29751}}}{1}
    {0x5f23a}{{\textcolor{red!80!black}{0x5f23a}}}{1}
    {0x4ae23}{{\textcolor{red!80!black}{0x4ae23}}}{1}
}

\lstdefinestyle{customasm3}{
    language=[x86masm]Assembler,
    basicstyle=\ttfamily\footnotesize\color{black},
    morekeywords={pushfq,cmp,je,jne,movl,callq,mov,push,xor,lea,nopl,and,shr},
    keywordstyle=\color{blue},
    identifierstyle=\color{red},
    commentstyle=\itshape\color{green!60!black},
    numbers=left,
    numberstyle=\tiny\color{gray},
    numbersep=8pt,
    tabsize=4,
    columns=fullflexible,
    keepspaces=true,
    showstringspaces=false,
    literate={806d}{{\textcolor{black}{806d}}}{1}
}

\newcommand{\rrdb}[1]{\multicolumn{1}{r}{\cellcolor{gray!30}\textbf{#1}}}
\newcommand{\crdb}[1]{\multicolumn{1}{c}{\cellcolor{gray!30}\textbf{\ensuremath{\boldsymbol{#1}}}}}

\definecolor{highlight}{rgb}{1, 1, 0}
\sethlcolor{highlight}

\begin{document}



\title{Fool Me If You Can: On the Robustness of Binary Code Similarity
Detection Models against Semantics-preserving Transformations}

\author{Jiyong Uhm}
\orcid{0009-0001-4955-662X}
\affiliation{%
  \institution{Sungkyunkwan University}
  \city{Suwon}
  \country{Republic of Korea}
}
\email{jiyong423@g.skku.edu}

\author{Minseok Kim}
\orcid{0009-0002-8993-3984}
\affiliation{%
  \institution{Sungkyunkwan University}
  \city{Suwon}
  \country{Republic of Korea}
}
\email{for8821@g.skku.edu}

\author{Michalis Polychronakis}
\orcid{0000-0002-3106-0343}
\affiliation{%
  \institution{Stony Brook University}
  \city{Stony Brook}
  \country{USA}
}
\email{mikepo@cs.stonybrook.edu}

\author{Hyungjoon Koo}
\orcid{0000-0003-0799-0230}
\authornote{Corresponding author.}
\affiliation{%
  \institution{Sungkyunkwan University}
  \city{Suwon}
  \country{Republic of Korea}
}
\email{kevin.koo@skku.edu}


\settopmatter{printacmref=false, printccs=false, printfolios=true}
\begin{abstract}
%
%
Binary code analysis plays an essential role
in cybersecurity, facilitating
reverse engineering to reveal
the inner workings of programs
in the absence of source code.
Traditional approaches, 
such as static and dynamic analysis,
extract valuable insights from 
stripped binaries, but often
demand substantial expertise and manual effort.
Recent advances in deep learning have opened
promising opportunities to enhance binary 
analysis by capturing 
latent features and disclosing
underlying code semantics.
Despite the growing number of 
binary analysis models based on machine learning, 
their robustness to adversarial 
code transformations at the binary level
remains underexplored to date. 
In this work, we evaluate the robustness of 
deep learning models for the task of 
binary code 
similarity detection (BCSD) under
semantics-preserving transformations.
%
The unique nature of machine 
instructions presents distinct challenges
compared to the typical input perturbations 
found in other domains. 
To achieve our goal, we introduce \sys, 
a system that evaluates the resilience of 
BCSD models using a diverse set of adversarial 
code transformations that preserve 
functional semantics.
We construct a dataset of 9,565 binary variants 
from 620 baseline samples by applying eight 
semantics-preserving transformations  
across six representative BCSD models.
Our major findings highlight several 
key insights:
i)~model robustness highly relies
on the design of the processing pipeline, 
including code pre-processing, 
model architecture, and internal 
feature selection, which collectively 
determine how code semantics 
are captured;
ii)~the effectiveness of adversarial 
transformations is bounded by 
a transformation budget, shaped by 
model-specific constraints 
such as input size limits and 
the expressive capacity of 
semantically equivalent instructions;
iii)~well-crafted adversarial 
transformations can be highly 
effective, even when introducing
minimal perturbations; and
iv)~such transformations 
efficiently disrupt the 
model’s decision (\eg misleading to
false positives or false negatives)
by focusing on
semantically significant instructions.

\end{abstract}
\maketitle


\section{Introduction}
Executable binaries (hereinafter referred to as binaries) are 
pervasive in modern computing, powering
personal computers, mobile devices, Internet of Things (IoT) systems, 
servers, cloud infrastructures, and smart appliances. 
The core of a binary is a sequence of machine-executable 
instructions derived from human-written source code, 
defining the operational logic of a computing device.
However, the form and content of a binary can vary widely, 
even when compiled from identical source code.
Such variations arise due to differences in
hardware architectures,
platforms,
file formats, optimization levels,
and compilation toolchains. 
%

Binary analysis plays a crucial role in security by uncovering
low-level code behavior
without access to its source code
(\ie reverse engineering).
It supports a wide range of applications, including
digital forensics~\cite{qi2022logicmem} and
analysis of proprietary protocols~\cite{caballero2007polyglot}.
This task becomes challenging when dealing 
with stripped binaries that lack high-level metadata 
such as function names, variable types, or 
structural annotations due to compiler optimizations 
and obfuscation techniques.
To extract semantic information, researchers rely on 
static analysis~\cite{yara, ida_pro, ghidra, binary_ninja}, 
dynamic analysis~\cite{egele2014blanket, afl, yun2018qsym}, 
or hybrid approaches~\cite{rawat2017vuzzer, shoshitaishvili2016state}.
Static analysis examines the internal structure of a binary (\eg 
symbol tables, disassembly, or decompiled code) without executing it,
whereas dynamic analysis observes runtime behavior, 
including system interactions such as
file operations, memory access patterns, and network activity.
Despite their effectiveness, these conventional approaches often
require domain expertise and laborious effort, 
limiting scalability and accessibility.
%

Recent advances in deep learning have opened
promising directions for enhancing binary analysis 
by enabling the extraction of 
latent features in high-dimensional spaces. 
These capabilities unlock new opportunities 
across several domains: \WC{1}~characterizing 
binary properties, such as programmer 
authorship~\cite{abuhamad2018large,
caliskan2015anonymizing,li2022ropgen}, 
compiler toolchain provenance~\cite{
du2023improving,
he2022binprov}, 
and malware detection~\cite{kruegel2006polymorphic,
chen2023continuous};
\WC{2}~inferring code semantics, including
plagiarism detection~\cite{luo2014semantics}, 
binary code similarity detection (BCSD)~\cite{xu2017neural,feng2016scalable,
ding2019asm2vec,massarelli2019safe,
zuo2018neural,
pei2020trex,
ahn2022practical,
yu2020order,
wang2024improving}, malware 
classification~\cite{downing2021deepreflect, 
someya2023fcgat,
lucas2023adversarial},
and vulnerability detection~\cite{pewny2015cross,
eschweiler2016discovre,
feng2016scalable,chandramohan2016bingo,
xu2017neural,
luo2023vulhawk}; 
and \WC{3}~supporting binary reversing tasks, 
such as
function name prediction~\cite{kim2023transformer,
he2018debin}, 
variable name prediction~\cite{chen2022augmenting,
he2018debin}, and type prediction~\cite{chen2022augmenting,
he2018debin}.
Emerging trends in artificial intelligence 
suggest that Machine-Learning-as-a-Service (MLaaS) 
will increasingly be adopted as a black-box oracle
in security contexts.
The broad spectrum of models
relies
on diverse features, including 
control flow graph (CFG) information, instruction counts, 
numerical constants, and instruction embeddings,
tailored to their specific analytical objectives.
Despite the proliferation of  
deep-learning-based models for binary analysis,
relatively few studies have investigated their robustness.
Early works~\cite{jia2022funcfooler, capozzi2023adversarial} 
explore adversarial attacks against 
BCSD models, but limit their scope
to a single type of semantics-preserving code transformation 
or to a narrow set of models.

In this paper, we focus on the models that leverage
code semantics to detect the similarity 
between binary code snippets for two key reasons. 
First, BCSD is a well-established 
task~\cite{haq2021survey,kim2022revisiting} 
with extensive research grounded in deep
neural networks~\cite{ahn2022practical,xu2017neural,
feng2016scalable,massarelli2019safe,pei2020trex,
ding2019asm2vec}.
Second, BCSD serves as an effective benchmark
for robustness evaluation due to the abundance 
of publicly available datasets and diverse code representations.
More importantly, our study is task-agnostic and generalizable
to other (aforementioned) applications.
To this end, we introduce \sys, a system for evaluating 
the robustness of BCSD models under 
a wide spectrum of binary code transformations.
Unlike input perturbations in computer 
vision~\cite{carlini2017towards,goodfellow2014explaining,
madry2017towards,szegedy2013intriguing}
or natural language processing~\cite{jin2020bert,
li2020bert,zou2023universal},
machine instructions pose distinct challenges
due to their unique and rich expressiveness.
For example, both instructions 
\cc{mov rax, 0} and \cc{xor rax, rax} 
set the \cc{rax} register to $0$, but differ in syntax
and potential side effects, highlighting the subtleties 
of semantic equivalence in binary code.
Such transformations, while preserving program semantics,
have long been studied in the context of
code diversification~\cite{koo2016juggling,koo2018compiler,
pappas2012smashing} and code 
obfuscation~\cite{junod2015obfuscator, tigress}.
Another challenge arises from
the variation in preprocessing
across different models, which affects
how raw machine instructions are
normalized, structured, and fed into the
learning pipeline.
In this work, we apply eight types of
semantics-preserving transformations~\cite{pappas2012smashing,
koo2018compiler,junod2015obfuscator,lucas2021malware}
across six representative BCSD models~\cite{ahn2022practical,
xu2017neural,feng2016scalable,massarelli2019safe,pei2020trex,ding2019asm2vec} 
to assess their robustness.

We construct a corpus of 9,565 variants (\ie transformed binaries)
from the 620 baseline samples by applying 
various semantics-preserving code transformations.
To comprehensively assess the robustness of BCSD models, 
we conducted several types of 
evaluation experiments: 
\WC{1} false negative (FN; incorrectly classifying 
similar code pairs as dissimilar) 
triggering transformations using
code diversification and 
obfuscation techniques,
\WC{2} false positive (FP; inaccurately classifying 
dissimilar code pairs as similar) 
triggering transformations, and
\WC{3} the transferability of the FP-triggering transformations across models.

Our extensive experiments yield noteworthy key findings.
First, model robustness relies heavily on the design of
the training pipeline, including code pre-processing steps, 
architectural choices, and internal feature representations.
These components collectively shape how code semantics
are captured in the embedding space.
For example, BinShot~\cite{ahn2022practical}, which 
does not incorporate CFG information, is susceptible to 
basic block reordering.
In contrast, Gemini~\cite{xu2017neural} and Genius~\cite{feng2016scalable},
both of which use CFG information, exhibit greater resilience to such transformations.
Second, the effectiveness of an adversarial transformation 
is inherently bounded by a transformation budget,
determined by model-specific constraints such as 
input size limits and the representational capacity (\ie expressivity)
of semantically equivalent instruction variants.
Third, well-engineered adversarial transformations
can achieve high attack success rates 
against target models with minimal perturbations.
For instance, our FP-triggering transformation 
reaches up to a 100\% success rate 
while introducing only 14.75 
additional instructions on average.
Lastly, using two explainable AI techniques
(\ie SHAP~\cite{lundberg2017unified},  
saliency map~\cite{saliency}),
we empirically discover that 
such transformations disrupt
the model’s decision by
distorting
internal attention patterns 
(\eg token importance).
%


This paper makes the following main contributions:
\begin{itemize}[leftmargin=*]
    %
    \item We introduce an adversarial perturbation based on greedy sampling 
    that induces false negatives and 
    false positives by
    leveraging eight 
    semantics-preserving code transformations.
    %
    \item We construct 9,565 adversarial binary variants
    by applying a range of 
    transformations designed to
    induce FNs by a BCSD model.
    \item We conduct 
    comprehensive evaluations on
    six state-of-the-art BCSD models
    to assess their robustness against
    both FN- and FP-triggering perturbations.
    \item We uncover key insights that inform 
    the development of more robust 
    BCSD models, particularly in the context 
    of adversarial resilience.
    %
\end{itemize}

\section{Background}
\label{ss:background}

%


\subsection{Semantics-preserving Code Transformations}
Despite their overlapping nature, 
we categorize code transformations 
into two groups: code diversification techniques, 
which aim to modify the code's footprint in memory
(\eg to defend against code reuse attacks),
and obfuscation techniques,
which aim to
deliberately confuse or obscure the code's functionality
(\eg to hinder reverse engineering).
The former typically
introduce no new or very few instructions,
while the latter often involve
the insertion of additional code,
such as inserting junk code 
or manipulating the CFG.

\PP{Code Obfuscation}
%
Code obfuscation can serve both benign
(\eg protecting intellectual property) 
and malicious (\eg evading malware 
from detection) purposes.
A wide range of tactics have been introduced 
to obscure code, including the insertion of 
unreachable or redundant instructions, 
CFG complications, 
and data structure modifications.
%
%
Obfuscation can be 
applied to various stages of
software development, including
at the source level 
(\eg Tigress~\cite{tigress}),
during compilation (\eg Obfuscator-LLVM~\cite{junod2015obfuscator}), 
or at the binary level (\eg VMProtect~\cite{vmprotect}, Thermida~\cite{thermida}, 
UPX~\cite{upx}).

\PP{Obfuscator-LLVM}
The LLVM compilation toolchain~\cite{LLVM} provides
a flexible and extensible framework based
on the concept of a \emph{pass}, \ie
modular units of 
analysis or transformation applied to the 
(LLVM) intermediate representation
during the compilation process.
Obfuscator-LLVM~\cite{junod2015obfuscator} 
implements a set of transformation passes to apply 
code obfuscation techniques at compilation time, including 
\WC{1} instruction substitution that replaces 
arithmetic operations with semantically equivalent 
alternatives, \WC{2} bogus control flow that inserts 
unreachable basic blocks to mislead static analysis, and 
\WC{3} control flow flattening that restructures a function's 
control flow into a single switch statement 
to obscure the original logic.
In this work,
we use Obfuscator-LLVM
as part of the tool set for
generating code variants.
%

\PP{Code Diversification} 
%
%
Code diversification is a technique 
that produces multiple variants of a binary, 
each preserving the original semantics while 
resulting in different memory layouts or execution paths.
%
%
On one hand, code diversification 
can defend against
code reuse attacks, such as 
return-oriented programming~\cite{roemer2012return,
prandini2012return}
and jump-oriented
programming~\cite{bletsch2011jump}, by 
invalidating the adversary's assumptions
about a program's memory layout.
%
%
On the other hand, malware authors can 
leverage code diversification
to evade signature-based detection 
or to modify static features 
to bypass machine-learning-based malware 
classifiers~\cite{lucas2021malware}.

\PP{In-place Code Randomization}
%
%
In-place code randomization~\cite{pappas2012smashing}
introduces binary-oriented instrumentation
with four transformation techniques 
of different spatial granularity 
(\eg instruction, basic block, function): 
\WC{1} \emph{instruction substitution} that replaces the original (and exploitable) instructions
(\ie gadget) with functionally equivalent ones;
\WC{2} \emph{intra-basic-block reordering} that 
relocates instructions in such a way that they 
are functionally identical within a basic block;
\WC{3} \emph{register preservation code reordering} that reorders the
push and pop instruction(s) from a function prologue and epilogue; and
\WC{4} \emph{register reassignment} that swaps registers under non-overlapping 
regions with a liveness
analysis of registers
(\ie tracing registers that hold active values during program execution).
%
%
In this work, we use
in-place code randomization~\cite{pappas2012smashing} 
as part of the tested code diversification techniques.

\PP{Compiler-assisted Code Randomization}
CCR~\cite{koo2018compiler} introduces
compiler-assisted binary instrumentation, 
which allows for 
fast and robust fine-grained code 
randomization at both the function and 
basic block levels.
It augments a binary
with a minimal set of 
transformation-assisting metadata
from the compiler toolchain.
We use 
\emph{inter-basic-block reordering} 
from CCR as part of our diversification transformations~\cite{koo2018compiler}.

\subsection{Binary Code Similarity Detection}
Binary code similarity detection
(BCSD) estimates the similarity 
of two or more (binary) code snippets in the absence of the corresponding source code. 
%
%
The main challenge of BCSD arises
from various compiler optimization pipelines that
produce semantically identical but different 
binaries. 
%
%
%
Numerous studies~\cite{haq2021survey} 
have focused on BCSD
due to the broad spectrum of its applications,
including (but not limited to)
bug discovery~\cite{pewny2014leveraging,pewny2015cross,eschweiler2016discovre,
feng2016scalable,chandramohan2016bingo,huang2017binsequence,feng2017extracting,
gao2018vulseeker,xu2017neural,
liu2018alphadiff, luo2023vulhawk}, 
malware clustering~\cite{hu2009large,hu2013mutantx,kim2019binary}, 
malware detection~\cite{kruegel2006polymorphic,bruschi2006detecting,cesare2013control}, 
malware lineage analysis~\cite{lindorfer2012lines,
jang2013towards}, and 
code clone detection~\cite{luo2014semantics}.
Recent advances in deep learning
open new opportunities to tackle
BCSD~\cite{xu2017neural,feng2016scalable,liu2018alphadiff,ding2019asm2vec,massarelli2019safe,zuo2018neural,gao2018vulseeker}.
Using recent transformer
architectures has become a common
approach~\cite{pei2020trex, 
ahn2022practical, wang2022jtrans, 
yu2020order}.
In parallel, graph neural networks are
widely adopted for capturing
structural and semantic code
representations~\cite{he24codeisnot, 
jia24crossinlining}.
%
%
We choose BCSD models to assess
their resilience under a diverse set of
code transformations.

\section{Robustness of ML-based BCSD Models}
\label{problem_statement}

\PP{Code Features for BCSD Models}
%
To infer the code semantics from a binary, 
BCSD models leverage both structural 
and instruction-level features.
Structural features are extracted 
\WC{1} directly from the control flow graph
(CFG)~\cite{liu2018alphadiff};
\WC{2} by enhancing a CFG 
with semantic information, 
such as through an attributed 
CFG~\cite{feng2016scalable, xu2017neural, 
gao2018vulseeker, jia24crossinlining} or 
a semantic-oriented 
graph~\cite{he24codeisnot}; or
\WC{3} by approximating execution paths 
through techniques 
like random walks~\cite{ding2019asm2vec} 
or micro-traces~\cite{pei2020trex}.
Instruction-level features are 
usually captured using various 
embedding techniques, such as
Word2Vec~\cite{mikolov2013distributed}
(in InnerEye~\cite{zuo2018neural}),
Instruction2vec~\cite{lee2019instruction2vec} 
(in SAFE~\cite{massarelli2019safe}), or 
Asm2Vec~\cite{ding2019asm2vec}, which
encode instruction sequences into 
dense vector representations.
With the emergence of the transformer architecture~\cite{transformer},
the semantic understanding of
BCSD models~\cite{pei2020trex,ahn2022practical,wang2022jtrans,yu2020order} has been further enhanced
by incorporating 
positional encoding and the attention mechanisms, 
enabling effective modeling of contextual and structural relationships
in binary code.

\begin{figure*}[t!]
    \centering
    \resizebox{0.99\linewidth}{!}{
    \includegraphics{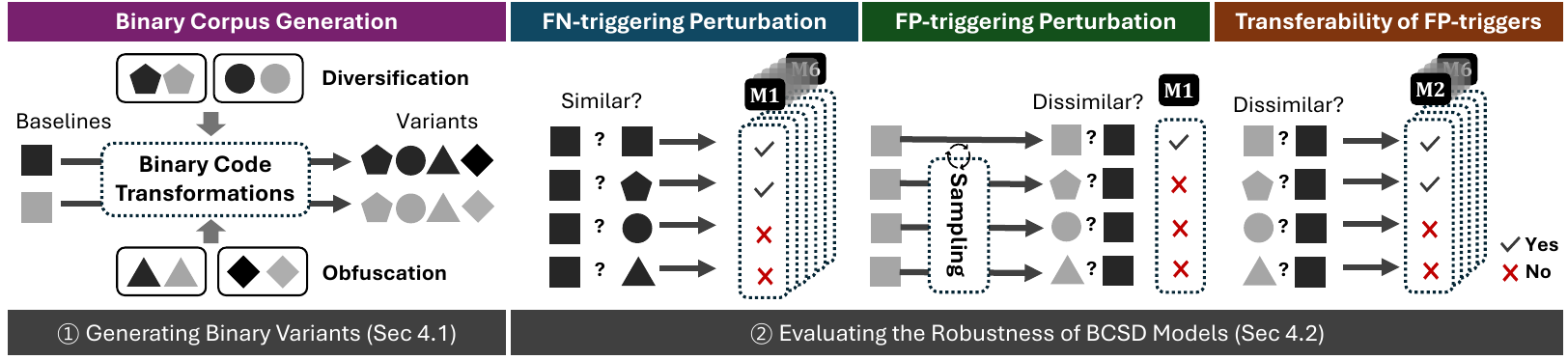}}
    \caption{
    Overview of the \sys system
    with two main components: \protect\WC{1} 
    generating binary variants
    with various semantics-preserving 
    code transformations (Section~\ref{ss:variants})
    and \protect\WC{2}
    evaluating the robustness
    of six pre-selected BCSD models
    (Section~\ref{ss:robustness}).
    %
    We assess the robustness of 
    the models using adversarial
    samples designed to trigger 
    either false negatives (FN) 
    or false positives (FP).
    Additionally, we investigate 
    the transferability of FP-trigger samples:
    \ie a sample that misleads
    one model affects others.
    Note that \cc{M1} to \cc{M6} denote 
    six BCSD models in our study.
    }
    \label{fig:overview}
\end{figure*}

\PP{Threat Model}
We assume an adversary, either 
curious or malicious, 
who can query a BCSD model 
(\ie black-box oracle)
where the model
takes a program binary (\eg in ELF 
format) as input. 
%
In this setting,
the model's architecture is 
\emph{unknown} and its internal 
parameters are \emph{fixed}; \ie
the attacker is disallowed to retrain 
the model or replace it 
with a different one of their choosing.
Instead, the adversary can manipulate 
the raw bytes of the binary 
(\ie via binary instrumentation) 
in an attempt to mislead the model's 
classification outcome.
The goal is to flip the prediction outcome, 
causing the model to inaccurately consider
similar code snippets as dissimilar (false negatives),
or dissimilar snippets as similar (false positives).
Importantly, these two attack types differ in both difficulty 
and adversarial requirements. 
Inducing false negatives is comparatively straightforward, 
as an attacker can evade detection by applying 
semantics-preserving obfuscations to malicious code. 
In contrast, inducing false positives is more demanding: 
the adversary must deliberately manipulate the model’s decision 
boundary to cause malicious code to be classified as benign, 
effectively steering the model’s internal representation. 
We present both perturbations in detail 
in Section~\ref{ss:robustness}.

\PP{Transformation Budget}
We adopt \emph{functions} as our 
transformation unit,
because they represent a logical part 
of code that encapsulates meaningful semantics.
This implies that we cannot arbitrarily
inject transformations across the entire 
binary, but are constrained within 
the boundaries of a given function.
Besides, existing BCSD models 
often impose constraints on the maximum 
number of instructions they can process 
at once, primarily due to architectural 
limitations of the underlying models.
For instance, four out of six models in 
our evaluation—Asm2Vec~\cite{ding2019asm2vec}, 
SAFE~\cite{massarelli2019safe}, 
Trex~\cite{pei2020trex}, and BinShot~\cite{ahn2022practical}, enforce
such restrictions, 
with respective caps of 500 
random walk instructions, 
512 instructions, 250 tokens, and 
256 tokens, respectively.
These differences stem from the internal
processing approaches of each model:
\WC{1}~BinShot~\cite{ahn2022practical} 
counts tokens,
each representing a single instruction; 
\WC{2}~Trex~\cite{pei2020trex} tokenizes
instructions into opcode and operand(s); and
\WC{3} Gemini~\cite{xu2017neural} 
and Genius~\cite{feng2016scalable} do
not impose any restrictions.
To ensure a fair comparison across models
with varying input constraints, 
we define a \emph{transformation budget},
which limits the number of instructions or  bytes
that may be added
through semantics-preserving 
code transformations.

\PP{Evaluation Challenges}
\label{ss:challenges}
Unlike natural language processing (NLP) models, 
generating semantics-preserving 
machine code fundamentally 
differs from producing semantically analogous text.
While human readers can often tolerate 
minor inconsistencies in language, 
CPUs interpret 
machine code with exact precision---even
a single-bit error can bring about 
critical failures or unintended behavior.
%
%
When the source code is available, 
binary instrumentation is relatively 
straightforward.
However, in practice, transformations must often 
be applied directly to compiled binaries, 
a process that is significantly more challenging.
Rewriting binaries without source code 
unavoidably requires extensive static 
analysis of assembly code, 
making the task complex and time-consuming.
Our experiments highlight this challenge:
applying a full suite of transformations
inspired by in-place code 
randomization~\cite{pappas2012smashing} failed 
on 23 out of 620 binaries, each exceeding 
our two-hour time limit
for transformation.
%
Although we impose practical constraints on generation time, 
a motivated attacker might allocate additional resources.
%
%
Moreover, implanting malicious code, such as 
reusing fragments from known malware, 
demands manual adjustments to 
control-flow-related 
features, as the injected code shifts 
instruction addresses and potentially 
disrupts the existing control flow.

\PP{Virtual Scenario}
%
Consider a scenario in which a website offers 
a BCSD service as an MLaaS instance
designed to detect malicious functionality 
by analyzing the uploaded executable. 
This online system incorporates 
a built-in BCSD model,
which returns binary classification decisions 
at the function granularity
regarding the presence of known 
adversarial content.
In this context, a curious adversary 
submits mutated versions of a seemingly 
benign executable, each embedding 
attacker-transformed code, intending to 
bypass the detection system.
To increase the chances of evasion, 
the adversary applies a range of 
semantics-preserving code transformations, 
crafting function-level variants that 
may deceive the BCSD model 
while retaining the malicious intent.
\section{\sys Design}
\label{s:trans}

\begin{figure}[t]
    \centering
    \resizebox{0.7\linewidth}{!}{
    \includegraphics[width=\linewidth,  trim=3 3 4 2.5, clip ]{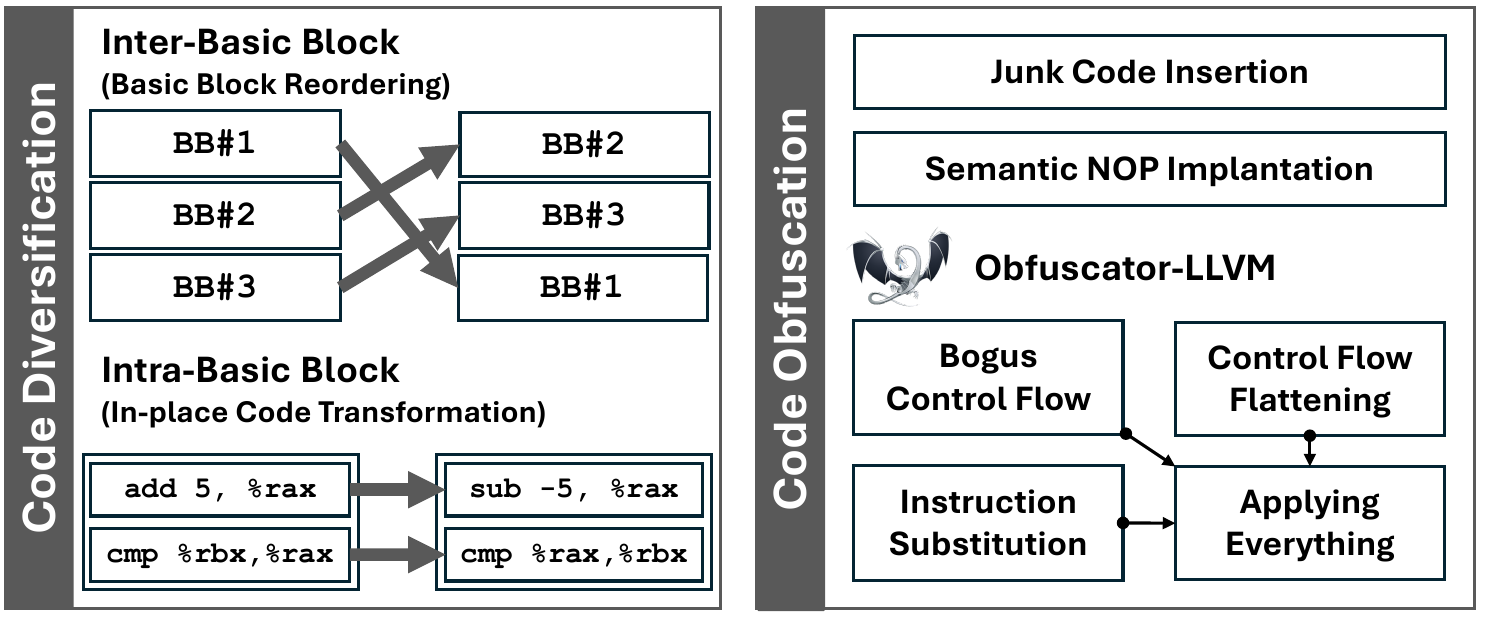}}
    \caption{Adversarial 
    semantics-preserving code 
    transformations in \sys. 
    We adopt a wide range of 
    transformation techniques from
    i)~code diversification 
    for defending against
    code reuse attacks (Section~\ref{sss:diversification}) and
    ii)~code obfuscation for
    making static analysis challenging 
   (Section~\ref{sss:obfuscation}).}
    \label{fig:transformations}
\end{figure}



%

\PP{\sys Overview}
\sys aims to evaluate
the robustness of BCSD models against 
a range of semantics-preserving code 
transformations.
%
%
%
\autoref{fig:overview}
presents an overview of \sys, 
illustrating both the generation 
of adversarial samples through 
semantics-preserving transformations 
and the subsequent robustness 
evaluation of BCSD models.
%
%
The transformation engine in \sys 
incorporates techniques from 
both code diversification and code obfuscation.
For diversification, it leverages 
in-place code transformation~\cite{pappas2012smashing} 
and inter-basic-block reordering~\cite{koo2018compiler}.
For obfuscation, it utilizes 
Obfuscator-LLVM~\cite{junod2015obfuscator}, 
semantic NOP implantation~\cite{lucas2021malware}, and 
junk code insertion, as depicted in~\autoref{fig:transformations}.
%
%

\subsection{Generating Binary Variants}
\label{ss:variants}

\subsubsection{Code Diversification Techniques}
\label{sss:diversification}
\,

\PP{In-place Code Randomization}
%
%
We adopt four techniques from in-place 
code randomization~\cite{pappas2012smashing}: 
instruction substitution,
intra-basic-block reordering, 
register preservation code reordering, and
register reassignment.
However we \emph{redesign} these
techniques for 64-bit ELF executables,
as the original implementation
has been limited to the 32-bit PE format.
First, we update all general-purpose registers to 
their 64-bit counterparts, such as
converting every 32-bit register to 
corresponding 64-bit register
(\eg \cc{eax} $\rightarrow$ \cc{rax})
and inserting the necessary opcode prefix byte for 
64-bit register operations
(\eg \cc{push rcx: 0x51}, \cc{push r8: \textbf{0x41} 0x51}).
Second, we enhance the register liveness analysis 
to more precisely track register usage under the
64-bit calling convention, 
which involves multiple argument-passing registers.
Third, we individually test each transformation 
on representative code samples to ensure 
correctness and maintain semantic equivalence.
In this paper, note that 
we use the terms ``randomization'' 
and ``transformation'' interchangeably.

\PP{Inter-Basic-Block Reordering}
The flexibility to reorder 
basic blocks is 
constrained by the operand size of 
control transfer instructions,
since CCR~\cite{koo2018compiler} enforces 
preservation of the original instruction size.
For instance, a single-byte operand in a jump 
or call instruction imposes 
a restricted displacement range, 
thereby constraining how far the target 
basic block can be relocated.
As a result, certain functions 
become ineligible 
for inter-basic-block reordering due to 
these distance limitations.
We identify and exclude such functions 
from code transformation.

\subsubsection{Code Obfuscation Techniques}
\label{sss:obfuscation}
\,

\PP{Semantic NOP Implantation}
A semantic NOP refers to a sequence of 
inserted instructions that preserve 
the original program behavior, leaving 
the surrounding code context unaffected.
Inspired by the 
context-free grammar approach 
proposed by Lucas \etal~\cite{lucas2021malware}, 
we generate semantic NOP sequences 
(within a predefined transformation budget)
via a two-step process:
\WC{1}~constructing a derivation tree 
based on the grammar’s transition rules, and
\WC{2}~traversing the tree to produce 
a variety of semantic NOP instruction sequences.
However, we observe that the original 
approach can bring about an infeasibly 
large search space under certain conditions.
For example, a register-specific state 
$S_r$ tracks a register 
pushed to the stack (\eg \cc{push rax} 
$\cdot$ $S_{rax}$ $\cdot$ \cc{pop rax}).
Then, there are two ways to generate the 
non-terminating state $S_{ef,r}$
where $S_{ef,r}$ and $S_{rf,r}$ 
represent equivalent states 
except the former applies to 
32-bit \cc{EFLAGS} and the latter
to 64-bit \cc{RFLAGS}: \ie
$S_{ef}\rightarrow\cc{push r}\cdot S_{ef,r}\cdot \cc{pop r}$ and
$S_{r}\rightarrow\cc{pushfd}\cdot S_{ef,r}\cdot \cc{popfd}$.
The second transition rule requires 
maintaining a separate state $S_{r}$ 
for each general-purpose register, 
significantly expanding the exploration space.
%
%
To mitigate this, as
in~\autoref{tbl-context-free-grammar-appendix},
we remove the 
register-specific state $S_r$
and adjust other states accordingly, 
preserving expressiveness while
improving efficiency in practice.
\begin{table*}[!t]\centering
\caption{Context-free grammar 
for our semantic NOP generation. 
The left column 
represents the original 
version proposed by Lucas 
\etal~\cite{lucas2021malware}, whereas the 
center represents the one that 
reduces the exploration space: \ie
the $S_r$ state (in red) has been eliminated,
updating other states (in blue) accordingly.
The right column
shows the legend
and x86 instructions.
%
}
\resizebox{1.0\linewidth}{!}{
    {\setlength\tabcolsep{4pt}
    \begin{tabular}{>{$}l<{$}>{$}c<{$}>{$}l<{$}>{$}l<{$}>{$}c<{$}>{$}l<{$}lcl}
    \toprule
      S &\rightarrow & Atom & S &\rightarrow & Atom&S & : & Starting symbol\\
  &| &S \cdot S &&| &S \cdot S&ef & : & \cc{EFLAGS}\\
  &| &\cc{bswap r}\cdot S\cdot \cc{bswap r}&&| &\cc{bswap r}\cdot S\cdot \cc{bswap r}&rf & : & \cc{RFLAGS}\\
  &| &\cc{xchg r1, r2}\cdot S\cdot \cc{xchg r1, r2}& &| &\cc{xchg r1, r2}\cdot S\cdot \cc{xchg r1, r2}&$\Phi$ & : & Empty string\\
  &| &\cc{push r}\cdot \textcolor{red}{S_{r}}\cdot \cc{pop r}& &| &\cc{push r}\cdot \textcolor{blue}{S}\cdot \cc{pop r}&\cc{arith} & : & Arithmetic operation (\eg add)\\
  &| &\textcolor{red}{\cc{pushfd}}\cdot \textcolor{red}{S_{ef}}\cdot \textcolor{red}{\cc{popfd}}&&| &\textcolor{blue}{\cc{pushfq}}\cdot \textcolor{blue}{S_{rf}}\cdot \textcolor{blue}{\cc{popfq}}&\cc{invarith} & : & Inverse of \cc{arith}\\

  Atom &\rightarrow & \cc{nop}&Atom &\rightarrow & \cc{nop}&\cc{logic} & : & Logical operation (\eg xor) \\
  &| &\cc{mov r, r}&&| &\cc{mov r, r}&r, r1, r2 & : & Register \\
  &| &\Phi &&| &\Phi&v & : & Random integer\\
  \textcolor{red}{S_{r}} &\textcolor{red}{\rightarrow} & \textcolor{red}{S}&&&&\\
\cmidrule{7-9}
  &\textcolor{red}{|} & \textcolor{red}{S_{r}\cdot S_{r}}&&&&\multicolumn{3}{c}{\textbf{x86-64 Instruction Set}}\\  
  &\textcolor{red}{|} & \textcolor{red}{\cc{pushfd} \cdot S_{ef,r} \cdot \cc{popfd}}&&&&\cc{bswap}&:&Reverses the byte order of registers\\

  \textcolor{red}{S_{ef}} &\rightarrow & S&\textcolor{blue}{S_{rf}} &\rightarrow & S&\cc{xchg}&:&Exchanges the values of two operands\\
  &| & \textcolor{red}{S_{ef}}\cdot \textcolor{red}{S_{ef}}&&| & \textcolor{blue}{S_{rf}}\cdot \textcolor{blue}{S_{rf}}&\cc{push}&:&Stores a value on the stack top\\
  &| & \cc{arith r, v}\cdot \textcolor{red}{S_{ef}}\cdot \cc{invarith r, v}&&| & \cc{arith r, v}\cdot \textcolor{blue}{S_{rf}}\cdot \cc{invarith r, v}&\cc{pop}&:&Retrieves a value from the stack top \\
  &| & \cc{push r}\cdot \textcolor{red}{S_{ef,r}}\cdot \cc{pop r}&&| & \cc{push r}\cdot \textcolor{blue}{S_{rf,r}}\cdot \cc{pop r} &\cc{pushfd/pushfq}&:&Stores the \cc{EFLAGS}/\cc{RFLAGS} register \\
 \textcolor{red}{S_{ef,r}} &\rightarrow & S&\textcolor{blue}{S_{rf,r}} &\rightarrow & S&&&onto the stack\\
  &\textcolor{red}{|} & \textcolor{red}{S_{r}}&&&&\cc{popfd/popfq}&:&Retrieves a value from the stack into\\
  &| & \textcolor{red}{S_{ef}}&&| & \textcolor{blue}{S_{rf}}&&&the \cc{EFLAGS}/\cc{RFLAGS} register\\
  &| & \textcolor{red}{S_{ef,r}}\cdot \textcolor{red}{S_{ef,r}}&&| & \textcolor{blue}{S_{rf,r}}\cdot \textcolor{blue}{S_{rf,r}}&\cc{mov}&:&Copies data from a source operand\\
  &| & \cc{arith r, v}\cdot \textcolor{red}{S_{ef,r}}&&| & \cc{arith r, v}\cdot \textcolor{blue}{S_{rf,r}}&&&to a destination operand\\
  &| & \cc{logic r, v}\cdot \textcolor{red}{S_{ef,r}}&&| &\cc{logic r, v}\cdot \textcolor{blue}{S_{rf,r}}&\cc{nop}&:&Performs no operation\\
  \bottomrule
    \end{tabular}
}

}
\label{tbl-context-free-grammar-appendix}
\end{table*}

\PP{Junk Code Insertion}
%
%
Junk code insertion introduces a sequence 
of unreachable instructions in the program.
In our setup, we intentionally 
place the junk code 
at the beginning of the function and 
prepend it with an unconditional jump 
to ensure it is always bypassed during execution.
This allows us 
to evaluate the impact of the length of 
dead code (\ie budget) on model behavior.
To further complicate BCSD models, 
we randomly sample unique instructions 
to fill the junk code, while ensuring 
the overall transformation remains 
within the defined transformation budget.

\PP{LLVM-based Obfuscation Techniques}
We extend our semantics-preserving 
code obfuscation 
in \sys to include source-level transformations, 
enabling the systematic application of 
more complex obfuscation techniques.
To this end, we integrate 
Obfuscator-LLVM~\cite{junod2015obfuscator}, 
a tool that applies obfuscation 
at the intermediate representation (IR) level
during the optimization phase of 
compilation. The applied transformations include
instruction substitution using 
arithmetic and logical operations, 
insertion of unreachable basic blocks, 
and control flow flattening.
These techniques are applied either individually 
or in combination.

\subsection{Evaluating the Robustness of BCSD Models}
\label{ss:robustness}

\begin{algorithm}[t]
    \caption{FP-triggering Instruction Sampling}
    \label{alg:opaquepredicate}

\begin{algorithmic}[1]
    \ttfamily
    \fontsize{7}{8}\selectfont
\STATE \textbf{Input:} corpus, target\_model, target\_func
\STATE \textbf{Output:} adv\_corpus
\STATE target\_dist $\leftarrow$ Get\_Distrib(target\_func)

\STATE adv\_corpus = []
\FOR{func in corpus}
    \STATE fp\_triggering\_codes = []
    \STATE fp\_triggering\_codes.push(UNCONDITIONAL\_JUMP)
    \WHILE{fp\_triggering\_codes.length < BUDGET}
        \STATE instr\_list  = []
        \WHILE{instr\_list.length < NUM\_CANDIDATES}
            \STATE instr\_list.push(Sample(target\_dist))
        \ENDWHILE
        \STATE scores = []
        \FOR{instr in instr\_list}
            \STATE attack\_seq $\leftarrow$ fp\_triggering\_codes + instr
            \STATE attack\_func  $\leftarrow$ attack\_seq + func
            \STATE score = target\_model(attack\_func, target\_func)
            \STATE scores.push(instr, score)
        \ENDFOR
        \STATE best\_instr, best\_score = Max(scores)
        \STATE fp\_triggering\_codes.push(best\_instr)
        \IF{best\_score > THRESHOLD}
            \STATE \textbf{break}
        \ENDIF
    \ENDWHILE
    \STATE adv\_corpus.push(fp\_triggering\_codes + func)
\ENDFOR
\STATE \RETURN adv\_corpus

\end{algorithmic}

\end{algorithm}

\subsubsection{FN-triggering Perturbation}
\label{sss:fnt-perturb}
Most adversarial samples 
with semantics-preserving 
transformations~(Section~\ref{ss:variants})
fall into the category of 
FN-triggering perturbations.
The goal of an FN-triggering perturbation
is to deceive the model into
incorrectly predicting that 
two semantically equivalent code snippets 
are dissimilar.

\subsubsection{FP-triggering Perturbation}
\label{sss:fpt-perturb}
In contrast to FN-triggering perturbations, 
generating adversarial samples that trigger 
false positives is non-trivial, as it requires 
deceiving the model into predicting 
that two dissimilar code snippets are similar.
To explore potential FP-triggering 
candidates, we design a greedy 
sampling strategy.
For this evaluation, we select ten functions 
from ten distinct binaries, each constrained 
by a fixed instruction budget (\autoref{tab:9_target}).
%

\PP{Greedy Sampling}
To trigger false positives in a model, 
we employ a greedy sampling strategy 
inspired by Zou \etal~\cite{zou2023universal}, which
demonstrates its effectiveness in generating 
adversarial suffixes against large language models.
Instead of using suffixes or perturbations 
at intermediate positions, we adopt 
adversarial prefixes to exploit the positional 
bias of language models, which tend to 
prioritize the beginning of the 
input context~\cite{lostmiddle}.
In essence, we first extract 
a target instruction distribution 
from a victim binary.
Next, we iteratively select instructions 
from this distribution when 
prepended to a target function, 
producing the most contextually 
similar result (in a greedy manner) 
relative to a chosen victim function.
This process continues until 
the transformation 
reaches a predefined budget, yielding a
sequence of instructions designed 
to trigger false positives.
%
%
%

\begin{figure}[t!]
    \begin{minipage}{.20\textwidth}
        \begin{lstlisting}[style=customasm, escapeinside={(*}{*)}]
    pushfq 
    sub    $0xc91a,%rcx
    (*\hl{push   \%r10}*)
    (*\hl{add    \$0x5c54,\%r10}*)
    sub    $0x9303,%r15
    add    $0xc866,%rbp
    xchg   %r13,%rbp
    xchg   %r13,%rbp
    sub    $0xc866,%rbp
    add    $0x9303,%r15
    (*\hl{pop    \%r10}*)
    add    $0xa1c3,%r14
    \end{lstlisting}
    \end{minipage}\hspace{2em}
    \begin{minipage}{.20\textwidth}
        \begin{lstlisting}[style=customasm2]
    push   %r9
    pushfq 
    nop
    sub    $0x5380,%rdi
    add    $0x5380,%rdi
    push   %r13
    pop    %r13
    popfq  
    pop    %r9
    sub    $0xa1c3,%r14
    add    $0xc91a,%rcx
    popfq
    \end{lstlisting}
    \end{minipage}
    \caption{Example of a semantic NOP sequence with the context-free grammar
    from Lucas \etal~\cite{lucas2021malware}. 
    The chunk of instructions would not 
    impact the semantics of a program 
    as a subsequent instruction(s) counteract
    the side effects of one or more instructions ahead.
    %
    %
    For example, pushing the value of \cc{r10} 
    to the stack (Line 3) and adding an
    arbitrary value (Line 4) 
    can be reversed by \cc{pop r10} (Line 11).}
    \label{lst:semanticnop}
\end{figure}

\begin{figure}[t!]
\begin{center}
\begin{minipage}{.35\textwidth}
        \begin{lstlisting}[style=customasm, escapeinside={(*}{*)}]
0000000000008050 <initialize_eval>:
    8050:    cmp    %rax,%rax
    8053:    je     0x830c
    (*\hl{8059:    movl   \$0x0,0xf21c9(\%rip)}*)        
    (*\hl{8063:    movl   \$0x0,0xf21c9(\%rip)}*)       
    (*\hl{806d:    callq  0x7fe0}*) 
    (*\hl{8072:    callq  0x7fe5}*)
    (*\hl{8077:    mov    \%ecx,0xf26bf(\%rip)}*)        
    (*\hl{807d:    jne    0x8090}*)
    ...
    830c:    push   %rbp
    830d:    push   %r15
    830f:    push   %r14
    \end{lstlisting}
    \end{minipage}\hspace{2em}
    \begin{minipage}{.35\textwidth}
        \begin{lstlisting}[style=customasm2]
    8311:    push   %r13
    8313:    push   %r12
    8315:    push   %rbx
    8316:    xor    %eax,%eax
    8318:    lea    0x29751(%rip),%rcx
    831f:    lea    0x5f23a(%rip),%rdx
    8326:    lea    0x4ae23(%rip),%rsi
    832d:    nopl   (%rax)
    8330:    mov    %eax,%edi
    8332:    and    $0x7,%edi
    8335:    mov    %eax,%r8d
    8338:    shr    $0x3,%r8d
    ...
    \end{lstlisting}
    \end{minipage}
    \end{center}
    \caption{
    Example of an FP-triggering perturbation 
    applied to the \cc{initialize\_eval} 
    function 
    from the \cc{sjeng} binary in SPEC2006.
    The instructions at addresses \cc{0x8059} 
    to \cc{0x807d} (Lines 4–9) represent 
    the adversarial sequence inserted into 
    the function prologue.
    To preserve original semantics, 
    the instructions
    at \cc{0x8050} and 
    \cc{0x8053} redirect the control flow 
    to skip over the injected code.
    %
    Note that we insert NOP padding as a length budget 
    at the function prologue 
    (\eg addresses between \cc{0x8050} 
    and \cc{0x830c} in this example), 
    followed by injecting the FP-triggering code 
    to enable further variations.
    Thus, the space between \cc{jne 0x8090} and 
    \cc{push \%rbp} is filled with NOP instructions, keeping
    the original semantics intact.
    }
\label{fig:fptriggerexample}
\end{figure}

\PP{Sampling Algorithm for FP-triggering Instructions}
%
%
Algorithm~\ref{alg:opaquepredicate} outlines 
the instruction sampling procedure 
in our FP-triggering perturbation transformation.
Given a specific BCSD model 
(\cc{target\_model}), the algorithm 
aims to transform a function 
from our corpus (\cc{corpus}) 
so that it appears similar to 
a chosen function (\cc{target\_func}).
We insert an unconditional 
jump instruction (\cc{UNCONDITIONAL\_JUMP}) 
at the start of the FP-triggering code sequence 
(\cc{fp\_triggering\_codes}) to ensure 
the injected instructions are never executed. 
Note that the (candidate) instructions 
(\eg \cc{NUM\_CANDIDATES} = 20) are extracted
from the target distribution (Lines 10–11) and 
the algorithm
iteratively selects the best-performing 
candidate (Line 20) and appends it to 
the sequence (Line 21).
The sampling process continues until 
the sequence reaches the transformation 
budget (Line 8) or the similarity score 
surpasses a pre-defined threshold (Line 22), 
indicating that the target model would 
classify the modified and target functions as similar.
Once the FP-triggering sequence is finalized, 
it is inserted into the target function, 
and the procedure moves on to the 
next function (Line 26).

\subsubsection{Transferability of FP-Triggers}
Finding a sample that triggers a 
false positive could be
very useful for an adversary because it would
help circumvent defenses based on BCSD models, as in the 
virtual scenario described in Section~\ref{problem_statement}.
Hence, we further investigate the transferability
of FP-triggers on other BCSD models.

\subsection{Code Perturbation Examples}
\autoref{lst:semanticnop} presents an example 
of a semantic NOP sequence 
consisting of $100$ bytes that 
introduces no functional side effects: \ie
the effect of each instruction is neutralized 
by a corresponding subsequent instruction.
For example, \cc{sub \$0xc91a,} 
$\%$\cc{rcx} on Line 2 is reversed by 
\cc{add \$0xc91a,} $\%$\cc{rcx} on Line 23,
maintaining the original register state.
\autoref{fig:fptriggerexample} illustrates a 
concrete example of an 
FP-triggering perturbation 
inserted into a function prologue.
Akin to semantic NOP sequences, 
this perturbation preserves 
the original program behavior.
In particular, the instructions 
between \cc{0x8050} and \cc{0x8053} 
(\ie \cc{UNCONDITIONAL\_JUMP}) 
maintain semantic correctness 
by rerouting the control flow 
around the inserted code.
The code snippets from 
\cc{0x8059} to \cc{0x807d} 
contain a sampled 
adversarial instruction pattern 
that induces a false positive 
in the model.
Notably, the address gap 
between \cc{0x807d} and \cc{0x830c} 
is padded with NOP instructions
in our experiments, 
enabling flexible accommodation 
of perturbations of varying 
sizes across different functions.

\section{Implementation}
\PP{Generating Binary Variants}
%
First, for in-place code randomization, 
the original implementation by Pappas
\etal~\cite{pappas2012smashing} was written
in Python~2 and designed for 
the PE format~\cite{peformat}.
To support 64-bit 
ELF~\cite{specification1993tool} binaries, 
we re-implemented the tool in Python~3, 
utilizing \cc{capstone}~\cite{capstone} 
for disassembly
and \cc{lief}~\cite{lief} for generating
binary variants.
%
Due to the internal complexity of transformations that require
liveness analysis, we impose 
a two-hour time limit per sub-transformation.
In total, we spent approximately 624 
hours (roughly one month) to generate 597 mutations.
Second, for semantic NOP generation, we 
adopt \cc{NLTK} (Natural Language 
ToolKit)~\cite{nltk},
a popular open-source library for 
natural language processing.
Based on our pre-defined context-free grammar, 
we modify the sequence generation function 
to randomly explore valid combinations, 
ultimately generating 500 unique semantic 
NOP sequences ranging from 20 to 100 bytes 
in length.
Third, to implant semantic NOPs and junk code, 
we first allocate a fixed-size placeholder 
using NOP bytes (\cc{0x90}), leveraging 
the \cc{-fpatchable-function-entry} 
compiler flag available in both GCC~\cite{GCC} 
and Clang~\cite{LLVM}. 
%
We overwrite this reserved space 
with our custom instruction sequences.

\PP{Model Evaluation}
We leverage a BCSD benchmark tool~\cite{marcelli2022machine} 
to evaluate various BCSD models~\cite{feng2016scalable,xu2017neural,massarelli2019safe,pei2020trex,ding2019asm2vec}
against the generated samples.
%
We made small modifications in restricting function pairs such that there is only a variation in transformation throughout our comparisons, and updating the metrics to support precision, recall, and F1 score (hereinafter F1).
%

\PP{Obfuscator-LLVM Reimplementation on LLVM 19}
We ported the LLVM-4-based Obfuscator-LLVM 
tool~\cite{junod2015obfuscator} to LLVM 19
(LLVM-19.1.4~\cite{llvm1914}), transitioning 
from the legacy pass manager to the new pass manager.
This upgrade ensures compatibility 
with other LLVM-19-based transformations 
we employ, namely, junk code insertion and 
semantic NOP implantation, which 
require support for the 
\cc{-fpatchable-function-entry} compiler flag.
Extensive changes in LLVM’s API over time
required significant modifications, 
as many functions 
in LLVM 4 have been deprecated 
or refactored in LLVM 19.
For instance, \cc{BinaryOperator::Create()} 
now requires an \cc{InsertPosition} instead of 
an \cc{Instruction}, and the method 
\cc{BinaryOperator::CreateFNeg()} has been 
eliminated~\cite{removecreatefneg}.
We address these alterations by consulting 
the LLVM source code, 
official documentation~\cite{LLVM}, 
and community discussions~\cite{LLVMdiscussion}, 
and by replacing deprecated constructs 
accordingly: \eg substituting \cc{CreateFNeg()} 
with \cc{UnaryOperator::CreateFNeg()}.

\section{Evaluating the Robustness of BCSD Models}

We evaluate the robustness of six BCSD models 
against eight semantics-preserving 
transformations.
%
%
%
We run our experiments on a 64-bit Ubuntu 20.04 system equipped with an Intel(R) Xeon(R) Gold 5218R CPU 3.00GHz, 512GB RAM, and two RTX A6000 GPUs.


\PP{Research Questions}
We define the following four
research questions to investigate
the impact of semantics-preserving transformations on
BCSD models.

\begin{itemize}[leftmargin=*]
    \item 
    \textbf{RQ1}: How do code diversification techniques trigger false negatives in BCSD models (Section~\ref{ss:RQ1})?
    \item 
    \textbf{RQ2}: How do code obfuscation techniques trigger false negatives in BCSD models (Section~\ref{ss:RQ2})?
    \item 
    \textbf{RQ3}: How well can we trigger false positives against a BCSD model (Section~\ref{ss:RQ3})?
    \item
    \textbf{RQ4}: How well can the FP-triggering adversarial samples be transferred to other BCSD models (Section~\ref{ss:RQ4})?
\end{itemize}

\PP{Binary Corpus}
\begin{table}[t!]
    \centering
     \caption{
Summary of our binary corpus. 
We generated a total of 9,565 adversarial variants 
based on 620 baseline binaries with the following
transformation techniques:
597 from in-place code transformation (ICT),
618 from inter-basic-block reordering (BBR),
3,020 from semantic NOP implantation (SNI) and 
junk code insertion (JCI), and
2,310 using Obfuscator-LLVM~\cite{junod2015obfuscator}).
The (*) marks indicate unsuccessful cases due to
exceeding the 2-hour processing limits in ICT 
(Section~\ref{ss:background});
mutating failures in BBR (\eg \cc{perlbench} 
and \cc{gobmk} under \cc{O1} optimization),
incompatible compiler versions (\eg \cc{findutils}), and
compilation failures in O-LLVM.
}
    \resizebox{0.75\columnwidth}{!}{
\begin{tabular}{llrr|rr|rrr|r}\toprule
\multirow{2}{*}{\textbf{Package}} &\multirow{2}{*}{\textbf{Version}} &\textbf{Default}&\multirow{2}{*}{\textbf{Functions}} &\multicolumn{2}{c}{\textbf{Diversification}} &\multicolumn{3}{c}{\textbf{Obfuscation}} &\multirow{2}{*}{\textbf{Total}}\\\cmidrule{5-9}
& & \textbf{Binaries}&&\textbf{ICT} &\textbf{BBR} &\textbf{SNI} &\textbf{JCI} &\textbf{O-LLVM} \\\midrule
\textbf{coreutils}~\cite{GNU} &8.2 &412 &37,823 &412 &412 &2,060 &2,060 &1,648 &6,592 \\
\textbf{binutils}~\cite{GNU} &2.27 &64 &97,893 &*63 &64 &320 &320 &*150 &917 \\
\textbf{findutils}~\cite{GNU} &4.6.0 &16 &3,683 &16 &16 &*0 &*0 &*0 &32 \\
\textbf{diffutils}~\cite{GNU} &3.2 &16 &1,740 &16 &16 &80 &80 &64 &256 \\
\textbf{nginx}~\cite{nginx}  &1.8.1 &4 &4,436 &*3 &4 &20 &20 &16 &63 \\
\textbf{putty}~\cite{putty} &0.66 &28 &25,505 &*14 &28 &140 &140 &112 &434 \\
\textbf{gzip}~\cite{gzip} &1.8 &4 &487 &4 &4 &20 &20 &16 &64 \\
\textbf{lighttpd}~\cite{lighttpd} &1.4.43 &4 &1,574 &4 &4 &20 &20 &16 &64 \\
\textbf{lvm2}~\cite{lvm2} &2.02.168 &8 &10,701 &*5 &8 &40 &40 &32 &125 \\
\textbf{vsftpd}~\cite{vsftpd} &3.0.3 &4 &2,248 &4 &4 &20 &20 &16 &64 \\
\textbf{miniweb}~\cite{miniweb} & N/A &4 &296 &4 &4 &20 &20 &16 &64 \\
\textbf{SPEC2006}~\cite{spec2006} & N/A &56 &40,491 &*52 &*54 &280 &280 &224 &890 \\
\midrule
\textbf{Total} & &620 &226,877 &597 &618 &3,020 &3,020 &2,310 &9,565 \\
\bottomrule
\end{tabular}
    }
    \label{tab:dataset_list}
\end{table}
We constructed a diverse set of baseline 
binaries from multiple sources, including 
essential system packages 
(\eg coreutils~\cite{GNU}, 
binutils~\cite{GNU}), 
popular open-source 
utilities (\eg nginx~\cite{nginx}, 
putty~\cite{putty}), and 
benchmark programs 
from SPEC2006~\cite{spec2006}.
We compile all executables with
Clang 
(19.1.4 for obfuscation and 
9.0.0 for diversification)
at four optimization levels 
(O0-O3), targeting 
the x86-64 ELF format.
First, we generate 1,215 adversarial 
samples via code diversification, 
as summarized in~\autoref{tab:dataset_list}, 
comprising 597 from in-place code 
transformation and 618 from 
inter-basic-block reordering.
Second, we produce 2,310 obfuscated variants 
using Obfuscator-LLVM. 
Compilation was successful for 
most packages, with the exception 
of some binutils and 
findutils binaries due to 
compiler version incompatibilities 
(Obfuscator-LLVM based on LLVM-19).
Third, we create 3,020 binaries through 
semantic NOP implantation and 
junk code insertion, with compilation 
failures only for findutils 
due to the same compatibility issue.
Lastly, we generate an additional 300 
mutations from SPEC2006 to evaluate 
the transferability of our FP-triggering transformation.
In total, our corpus consists of 10,185
binaries: 9,565 transformed variants 
and 620 original baseline binaries.

\PP{Target Models}
We evaluate six representative BCSD models:
Genius~\cite{feng2016scalable}, Gemini~\cite{xu2017neural}, Asm2Vec~\cite{ding2019asm2vec}, SAFE~\cite{massarelli2019safe},
Trex~\cite{pei2020trex}, and BinShot~\cite{ahn2022practical}.
It is noteworthy to mention that 
we carefully select different
types: two models~\cite{feng2016scalable,xu2017neural} based on graph neural networks,
two models~\cite{ding2019asm2vec,massarelli2019safe} using embeddings, 
and another two models~\cite{pei2020trex,ahn2022practical}
that adopt BERT-based language models.


\subsection{Impact of Code Diversification Techniques}
\label{ss:RQ1}


\begin{table*}[t!]\centering
\caption{(P)recision, (R)recall, and F1 of 
six BCSD Models against the adversarial 
samples with inter-basic-block reordering (BBR) and in-place code transformation (ICT). 
Each entry records the performance of 
``before $\rightarrow$ after transformation (difference)''.
The bold represents the highest 
performance drop
across different models for each transformation.
While we observe substantial drops in recall against ICT, 
most models remain robust against BBR with the exception
of BinShot (0.936 $\downarrow$).
Besides, the models that capture dynamic features 
(\eg Asm2Vec, Trex) exhibit less performance
degradation than others.
}
\label{tab:5_bbr_and_ipr}
\scriptsize
\resizebox{0.99\linewidth}{!}{

\begin{tabular}{ll|cccccc}\toprule
& &\textbf{Genius~\cite{feng2016scalable}} &\textbf{Gemini~\cite{xu2017neural}} &\textbf{Asm2Vec~\cite{ding2019asm2vec}} &\textbf{SAFE~\cite{massarelli2019safe}} &\textbf{Trex~\cite{pei2020trex}} &\textbf{BinShot~\cite{ahn2022practical}} \\\midrule
\multirow{3}{*}{\textbf{ICT}} &\textbf{P} &$0.835\rightarrow 0.793 \ (0.042\downarrow )$	 &$0.864\rightarrow 0.824 \ (0.040\downarrow )$	 &$0.890\rightarrow 0.807 \ (\textbf{0.083}\downarrow )$	 &$0.860\rightarrow 0.823 \ (0.037\downarrow )$	 &$0.925\rightarrow 0.929 \ (0.004\uparrow )$		 &$0.990\rightarrow 0.967 \ (0.023\downarrow )$	 \\
&\textbf{R} &$0.943\rightarrow 0.564 \ (0.379\downarrow )$	 &$0.886\rightarrow 0.554 \ (0.332\downarrow )$	 &$0.819\rightarrow 0.733 \ (0.086\downarrow )$	 &$0.782\rightarrow 0.509 \ (0.273\downarrow )$	 &$0.905\rightarrow 0.831 \ (0.074\downarrow )$	 &$0.953\rightarrow 0.271 \ (\textbf{0.682}\downarrow )$	 \\
&\textbf{F1} &$0.886\rightarrow 0.659 \ (0.227\downarrow )$	 &$0.875\rightarrow 0.663 \ (0.212\downarrow )$	 &$0.814\rightarrow 0.768 \ (0.046\downarrow )$	 &$0.819\rightarrow 0.629 \ (0.190\downarrow )$	 &$0.915\rightarrow 0.877 \ (0.038\downarrow )$	 &$0.971\rightarrow 0.423 \ (\textbf{0.548}\downarrow )$	 \\
\midrule
\multirow{3}{*}{\textbf{BBR}} &\textbf{P} &$0.839\rightarrow 0.831 \ (0.008\downarrow )$	 &$0.868\rightarrow 0.859 \ (0.009\downarrow )$	 &$0.688\rightarrow 0.685 \ (0.003\downarrow )$	 &$0.843\rightarrow 0.837 \ (0.006\downarrow )$	 &$0.930\rightarrow 0.927 \ (0.003\downarrow )$	 &$0.990\rightarrow 0.811 \ (\textbf{0.179}\downarrow )$	 \\
&\textbf{R} &$0.952\rightarrow 0.907 \ (0.045\downarrow )$	 &$0.905\rightarrow 0.857 \ (0.048\downarrow )$	 &$0.698\rightarrow 0.687 \ (0.011\downarrow )$	 &$0.762\rightarrow 0.695 \ (0.067\downarrow )$	 &$0.897\rightarrow 0.861 \ (0.036\downarrow )$	 &$0.972\rightarrow 0.036 \ (\textbf{0.936}\downarrow )$	 \\
&\textbf{F1} &$0.892\rightarrow 0.867 \ (0.025\downarrow )$	&$0.886\rightarrow 0.858 \ (0.028\downarrow )$	 &$0.693\rightarrow 0.686 \ (0.007\downarrow )$	 &$0.800\rightarrow 0.759 \ (0.041\downarrow )$	 &$0.913\rightarrow 0.893 \ (0.020\downarrow )$	 &$0.981\rightarrow 0.069 \ (\textbf{0.912}\downarrow )$	\\
\bottomrule
\end{tabular}

}
\end{table*}


\PP{In-place Code Transformation}
The first half of \autoref{tab:5_bbr_and_ipr} summarizes
the performance degradation after in-place code transformation (ICT).
We observe that the adversarial variants successfully trigger 
false negatives in a majority of models, significantly dropping recall.
However, Asm2Vec~\cite{ding2019asm2vec} and Trex~\cite{pei2020trex} demonstrate
a relative robustness against ICT.
BinShot~\cite{ahn2022practical} exhibits the lowest recall (0.271) against ICT
although it also adopts a BERT-based language model like Trex.
We hypothesize that BinShot relies on the correlation between instructions 
alone (for inference) without considering dynamic features 
like micro-traces~\cite{pei2020trex} in Trex.

\PP{Inter-Basic-Block Reordering}
Similarly, the second half of \autoref{tab:5_bbr_and_ipr} shows
the performance drops with inter-basic-block reordering (inter-BBR).
This led to a high false negative rate against BinShot~\cite{ahn2022practical}, 
reducing the recall score to as low as 0.036.
Except for BinShot, we observe the other models are robust against inter-BBR 
presumably because they take an attributed CFG (\eg Genius~\cite{feng2016scalable},
Gemini~\cite{xu2017neural}), a random walk (\eg Asm2Vec~\cite{ding2019asm2vec}),
or a micro-trace (\eg Trex~\cite{pei2020trex}) into account.
Meanwhile, SAFE~\cite{massarelli2019safe} inserts instruction embeddings into an attentive neural network,
indicating that the order of instructions is not essential. 

\subsection{Impact of Code Obfuscation Techniques}
\label{ss:RQ2}

\begin{figure*}[t]
    \centering
    \resizebox{0.99\linewidth}{!}{
    \includegraphics[width=\linewidth,  trim=8 17 6 7.5, clip ]{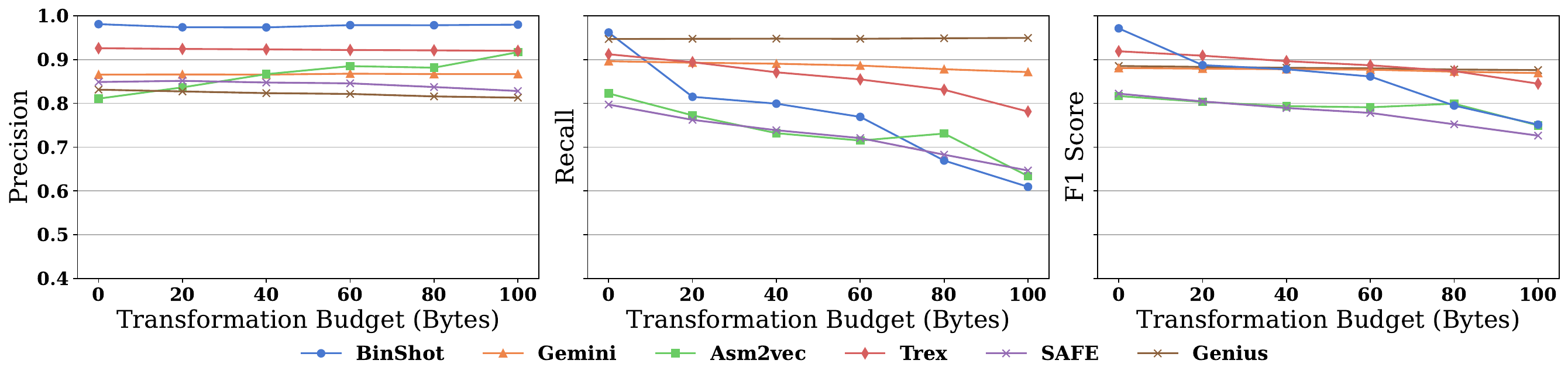}}
    \caption{Precision, Recall, and F1
    of various BCSD models with a given
    transformation budget (\eg 
    size of a semantic NOP in bytes).
    The models adopted a graph neural network 
    like Genius~\cite{feng2016scalable} and 
    Gemini~\cite{xu2017neural} tend to be 
    robust against semantic NOP implantation, 
    while the others do not (\eg significant drops in recall).
    We discuss precision with
    FP-triggering perturbation
    in Section~\ref{ss:RQ3}.
    Note that a byte does not necessarily 
    correspond to a single token or instruction; 
    \eg 
    our experiments insert
    around 7 and 25 instructions on average
    under the budgets of 20 and 100 bytes, respectively.
    }
    \label{fig:2_semanticnop}
\end{figure*}

\PP{Semantic NOP Implantation}
\autoref{fig:2_semanticnop} illustrates how the performance of each model
decreases after inserting a semantic NOP at a function entry.
With our transformation budget, we 
incrementally increase it by 20 bytes up to 100 bytes.
Each model exhibits varying drop rates; \eg
Asm2Vec~\cite{ding2019asm2vec}, SAFE~\cite{massarelli2019safe}, and 
BinShot~\cite{ahn2022practical} drastically drop in recall 
because they rely on instruction-level features.
Trex~\cite{pei2020trex} shows a moderate decline.
However, graph neural network-based models 
(Genius~\cite{feng2016scalable} and Gemini~\cite{xu2017neural}) 
are robust against semantic NOP insertion because it
impacts only the first basic block while maintaining a graph structure.
Note that the precision metrics are persistent regardless of the budget
because they pertain to false positives, which we cover in Section~\ref{ss:RQ3}.

\PP{Junk Code Insertion}
\begin{figure*}[t]
    \centering
    \resizebox{0.95\linewidth}{!}{
    \includegraphics[width=\linewidth,  trim=8 17 6 7.5, clip ]{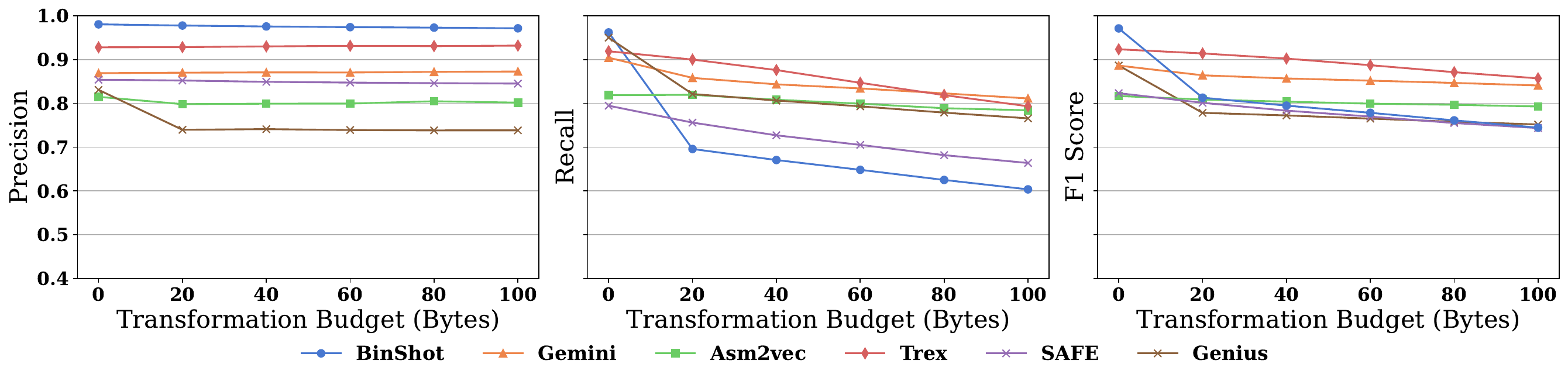}}
    \caption{Precision, Recall, and F1 of various BCSD Models in relation to the length of inserted junk code sequence.
    Notably, Asm2Vec~\cite{ding2019asm2vec} demonstrates substantial robustness, which can be attributed to the random walk that dynamically collects assembly sequences by traversing the assembly 
    instructions.
    On average, 6 and 16 instructions are inserted 
    under budget constraints of 20 and 100 bytes, 
    respectively.
    %
    }
    \label{fig:5_randomop}
\end{figure*}
As in \autoref{fig:5_randomop}, inserting 
junk code negatively affects the overall performance.
Gemini~\cite{xu2017neural}, Genius~\cite{feng2016scalable}, Asm2Vec~\cite{ding2019asm2vec}, 
and Trex~\cite{pei2020trex} demonstrate 
their robustness in recall compared to others.
Asm2Vec~\cite{ding2019asm2vec} displays the least impact on 
injecting arbitrary instructions, which we hypothesize
a random walk mitigates its negative effect 
by excluding the injected when walking a possible execution path.
It is noteworthy to mention that Gemini and Genius demonstrate
performance degradation against junk code insertion 
(where they do not against semantic NOP implantation) because
it introduces a new basic block with a jump instruction.

\PP{LLVM-based Obfuscation}
%
%
\autoref{tab:8_obfuscation} summarizes 
the impact of LLVM-based obfuscation on all models.
In general, we observe a universal 
decline in performance where
every model shows significant drops when applying 
all obfuscation techniques.
Unlike performance degradation against 
other semantics-preserving transformation
techniques, BinShot~\cite{ahn2022practical} 
exhibits remarkable robustness against 
LLVM-based obfuscation, which
implies that BinShot is capable of better 
inferring underlying code context.
In a similar vein, Asm2Vec~\cite{ding2019asm2vec} 
shows its relative robustness
against all LLVM-based code obfuscations.
Notably, Genius~\cite{feng2016scalable} and Gemini~\cite{xu2017neural}
display a significant reduction in performance against
control flow flattening and bogus control flow, which indicates
that a CFG perturbation method is effective.

\begin{table*}[t!]\centering
\caption{Performance of six BCSD models against various LLVM-based obfuscations, showing changes in the (P)recision, (R)ecall, and F1 scores compared to the baseline performance without obfuscation.
BCF, SUB, FLA, and ALL denote a bogus control flow, instruction substitution, control flow flattening, and a combination of all three, respectively.
The bold represents 
the most significant drop
across different models
for each obfuscation.
All models experience 
performance drops, with 
BinShot remaining relatively robust.
However, models that rely on dynamic features are particularly affected
from CFG perturbation (\eg Genius 0.673 $\downarrow$).
}
\label{tab:8_obfuscation}
\scriptsize
\resizebox{0.99\linewidth}{!}{

\begin{tabular}{l|r|rrrr|r|rrrr|r|rrrrr}\toprule
&\multicolumn{5}{c}{\textbf{Genius~\cite{feng2016scalable}}} &\multicolumn{5}{c}{\textbf{Gemini~\cite{xu2017neural}}} &\multicolumn{5}{c}{\textbf{Asm2Vec~\cite{ding2019asm2vec}}} \\
\cmidrule{3-6}
\cmidrule{8-11}
\cmidrule{13-16}
&\textbf{Base} &\textbf{BCF} &\textbf{SUB} &\textbf{FLA} &\textbf{ALL} &\textbf{Base} &\textbf{BCF} &\textbf{SUB} &\textbf{FLA} &\textbf{ALL} &\textbf{Base} &\textbf{BCF} &\textbf{SUB} &\textbf{FLA} &\textbf{ALL} \\
\cmidrule{3-6}
\cmidrule{8-11}
\cmidrule{13-16}
\textbf{P} &0.825 &$0.120\downarrow $ &$0.005\downarrow $ &\crdb{0.303 \downarrow}  &$0.308\downarrow $ &0.865 &$0.081\downarrow $ &$0.000\uparrow $ &$0.261\downarrow $ &\crdb{0.357 \downarrow} &0.818 &$0.009\uparrow$ &\crdb{0.007\downarrow} &$0.019\uparrow $ &$0.002\downarrow $ \\
\textbf{R} &0.952 &$0.363\downarrow $ &$0.002\downarrow $ &\crdb{0.673 \downarrow} &$0.680\downarrow $ &0.905 &$0.328\downarrow $ &$0.001\downarrow $ &$0.611\downarrow $ &\crdb{0.722 \downarrow} &0.788 &$0.113\downarrow $ &$0.006\downarrow $ &$0.193\downarrow $ &$0.395\downarrow $ \\
\textbf{F1} &0.884 &$0.242\downarrow $ &$0.004\downarrow $ &\crdb{0.520 \downarrow} &$0.527\downarrow $ &0.885 &$0.220\downarrow $ &$0.000\downarrow $ &$0.490\downarrow $ &\crdb{0.615 \downarrow} &0.803 &$0.059\downarrow $ &$0.006\downarrow $ &$0.107\downarrow $ &$0.272\downarrow $ \\
\toprule
&\multicolumn{5}{c}{\textbf{SAFE~\cite{massarelli2019safe}}} &\multicolumn{5}{c}{\textbf{Trex~\cite{pei2020trex}}} &\multicolumn{5}{c}{\textbf{BinShot~\cite{ahn2022practical}}} \\
\cmidrule{3-6}
\cmidrule{8-11}
\cmidrule{13-16}
&\textbf{Base} &\textbf{BCF} &\textbf{SUB} &\textbf{FLA} &\textbf{ALL} &\textbf{Base} &\textbf{BCF} &\textbf{SUB} &\textbf{FLA} &\textbf{ALL} &\textbf{Base} &\textbf{BCF} &\textbf{SUB} &\textbf{FLA} &\textbf{ALL} \\
\cmidrule{3-6}
\cmidrule{8-11}
\cmidrule{13-16}
\textbf{P} &0.850 &\crdb{0.126 \downarrow} &$0.003\downarrow $ &$0.203\downarrow $ &$0.299\downarrow $ &0.924 &$0.009\downarrow $ &$0.001\uparrow $ &$0.001\uparrow $ &$0.042\downarrow $ &0.979 &$0.012\downarrow $ &$0.001\downarrow $ &$0.008\downarrow $ &$0.021\downarrow $ \\
\textbf{R}  &0.803 &\crdb{0.364 \downarrow} &\crdb{0.018 \downarrow} &$0.494\downarrow $ &$0.566\downarrow $ &0.917 &$0.217\downarrow $ &$0.000\uparrow $ &$0.370\downarrow $ &$0.647\downarrow $ &0.961 &$0.211\downarrow $ &$0.015\downarrow $ &$0.214\downarrow $ &$0.402\downarrow $ \\
\textbf{F1} &0.826 &\crdb{0.279 \downarrow} &\crdb{0.011 \downarrow} &$0.407\downarrow $ &$0.494\downarrow $ &0.920 &$0.128\downarrow $ &$0.001\uparrow $ &$0.233\downarrow $ &$0.507\downarrow $ &0.970 &$0.125\downarrow $ &$0.008\downarrow $ &$0.125\downarrow $ &$0.264\downarrow $ \\
\bottomrule
\end{tabular}
}
\end{table*}

\subsection{FP-triggering Perturbation}
\label{ss:RQ3}
\begin{figure}[t]
    \centering
    \resizebox{0.6\columnwidth}{!}{
    \includegraphics[width=\columnwidth,  trim=8 10 13 6.5, clip ]{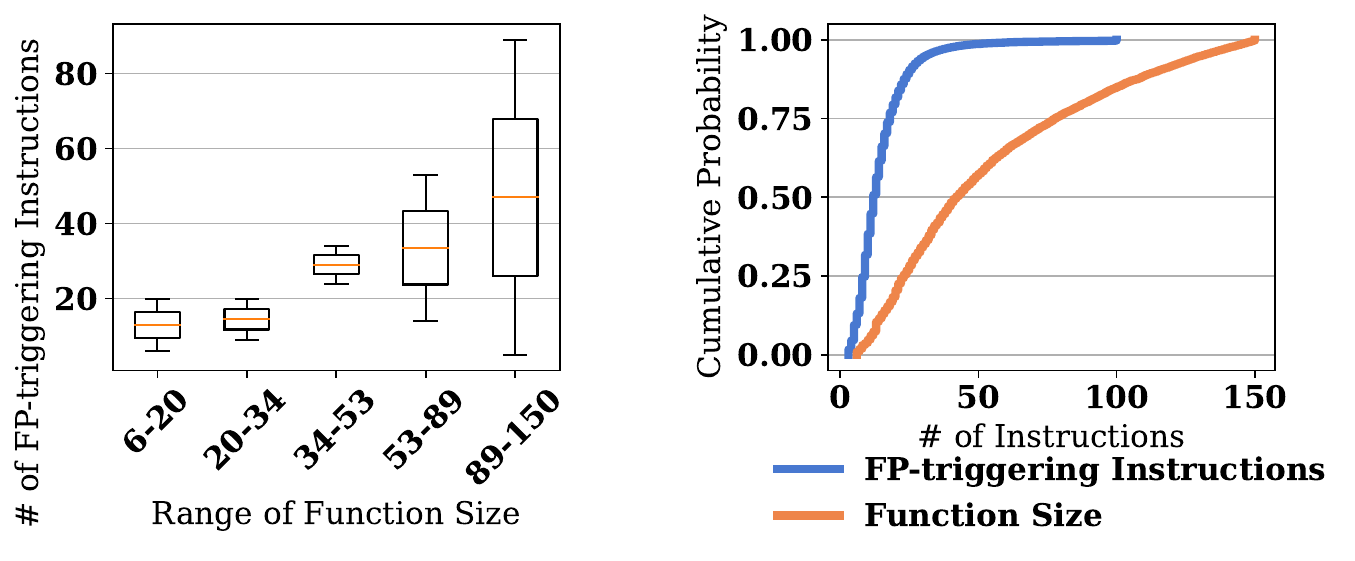}}
    \caption{The relation between the number of FP-triggering instructions and function size for successful attacks in the FP-triggering perturbation attack.
    The box plot (left) illustrates that the number of 
    FP-triggering instructions generally increases with function size, while
    the cumulative distribution function (right)
    indicates that most FP-triggering instructions
    involve fewer than 50 instructions.}
    \label{fig:9_cdf_box}
\end{figure}

We choose a target function from a binary, aiming to
construct a series of FP-triggering instructions that
mislead a model toward a false positive (Section~\ref{sss:fnt-perturb}).
The left side of 
\autoref{tab:9_target} illustrates the results of our 
perturbation attack against BinShot~\cite{ahn2022practical}.
Even with the minimum transformation budget of 20 instructions, 
we observe that the attack success rate (ASR) reaches to 98.3\%.
As the budget increases (up to 100 instructions), 
ASR has been close to almost 100\%.
%
%
Moreover, we investigate the average 
size of FP-triggering instructions
for a successful attack, which is approximately 57.58 bytes 
($\leq$ 50 bytes in most cases) or 
14.75 instructions on average
within the pre-defined transformation budget (Section~\ref{problem_statement}).
\autoref{fig:9_cdf_box} depicts the positive relationship,
in general, between FP-triggering code size and function size.


\subsection{Transferability of FP-trigger}
\label{ss:RQ4}
This section investigates 
the transferability of 
FP-triggering perturbations  
(originally crafted to target 
the BinShot~\cite{ahn2022practical} 
model) when applied to other BCSD models.
We use 10,366 functions from 300 
SPEC2006 binaries (within a defined 
transformation budget), aiming to mislead 
the models into judging that 
a target function from~\autoref{tab:9_target} 
is similar to a given (victim) function, 
despite their dissimilarity.
The right side of~\autoref{tab:9_target} 
reports the accuracy of 
other BCSD models under 
this perturbation attack.
%
Our findings indicate that the attack 
exhibits varying degrees of transferability 
across models: while it transfers effectively 
to certain architectures, 
others are relatively more resilient.
In particular, transferability is stronger 
among models that share similar 
instruction-level 
or token-level embeddings, while
CFG- or graph-centric GNN-based models 
demonstrate comparatively more robustness 
to such transfer.
%
Notably, Trex~\cite{pei2020trex}, 
a BERT-based architecture like BinShot, 
shows strong transferability 
between similar model types.

\begin{table*}[!t]\centering
\caption{The attack success rate (ASR) of the FP-triggering perturbation attack (left) and the accuracy score results of our transferability analysis (right). The attack aims to trigger false positives between the ten functions (victim functions) and 10,366 dissimilar functions. Bold numbers highlight the best attack results.
}
\label{tab:9_target}
\scriptsize
\resizebox{0.99\linewidth}{!}{

\begin{tabular}{llrrrrr|rrrrr}\toprule
\multirow{2}{*}{\textbf{Binary}} &\multirow{2}{*}{\textbf{Function}} &\multicolumn{5}{c|}{\textbf{Maximum Budget / ASR ($\uparrow$)}} &\multicolumn{5}{c}{\textbf{Target Model / Accuracy ($\downarrow$)}} \\\cmidrule{3-12}
& &\textbf{20} &\textbf{40} &\textbf{60} &\textbf{80} &\textbf{100} &\textbf{Genius} &\textbf{Gemini} &\textbf{Asm2Vec} &\textbf{SAFE} &\textbf{Trex} \\\midrule
bzip2 &\cc{BZ2\_bzBuffToBuffDecompress} &0.850 &0.981 &0.994 &0.998 &\rrdb{0.999} \vline &0.988 &0.996 &0.979 &0.974 &\rrdb{0.683} \\
gcc &\cc{optimize\_inline\_calls} &0.944 &0.994 &0.998 &\rrdb{0.999} &\rrdb{0.999} \vline &0.985 &0.992 &\rrdb{0.557} &0.992 &0.917 \\
gzip &\cc{zip} &0.900 &0.986 &0.995 &0.998 &\rrdb{0.999} \vline &\rrdb{0.942} &0.995 &0.979 &0.969 &0.997 \\
lighttpd &\cc{network\_write\_file\_chunk\_sendfile} &0.671 &0.971 &0.992 &0.995 &\rrdb{0.996} \vline &0.992 &0.989 &0.893 &0.948 &\rrdb{0.880} \\
lvm &\cc{lock\_vol} &0.787 &0.971 &0.986 &0.988 &\rrdb{0.989} \vline &0.975 &0.978 &0.996 &0.967 &\rrdb{0.869} \\
md5sum &\cc{md5\_stream} &0.879 &0.985 &0.992 &0.995 &\rrdb{0.996} \vline &0.974 &0.953 &\rrdb{0.878} &0.931 &0.898 \\
miniweb &\cc{\_mwProcessPost} &0.669 &0.951 &0.983 &0.990 &\rrdb{0.996} \vline &0.989 &0.993 &0.867 &0.973 &\rrdb{0.788} \\
nginx &\cc{ngx\_http\_create\_request} &0.611 &0.947 &0.981 &0.989 &\rrdb{0.993} \vline&0.967 &0.959 &0.863 &0.910 &\rrdb{0.773} \\
putty &\cc{new\_connection} &0.651 &0.959 &0.992 &0.998 &\rrdb{0.999} \vline &0.945 &0.993 &0.992 &0.987 &\rrdb{0.802} \\
vsftpd &\cc{vsf\_sysutil\_sendfile} &0.983 &\rrdb{1.000} &\rrdb{1.000} &\rrdb{1.000} &\rrdb{1.000} \vline &0.977 &\rrdb{0.764} &0.963 &0.981 &0.986 \\
\bottomrule
\end{tabular}
}
\end{table*}

\subsection{FP-Trigger Analysis with Explainable AI}
\label{ss:RQ5}
In this section, we extend our evaluation 
by analyzing model behavior 
using explainable AI (XAI) techniques.
We examine 
the FP-triggering perturbation input 
on BinShot~\cite{ahn2022practical} 
through SHAP~\cite{lundberg2017unified} 
and saliency analysis~\cite{saliency}.

\begin{figure}[t]
    \centering
    \resizebox{0.95\columnwidth}{!}{
    \includegraphics[width=\columnwidth]{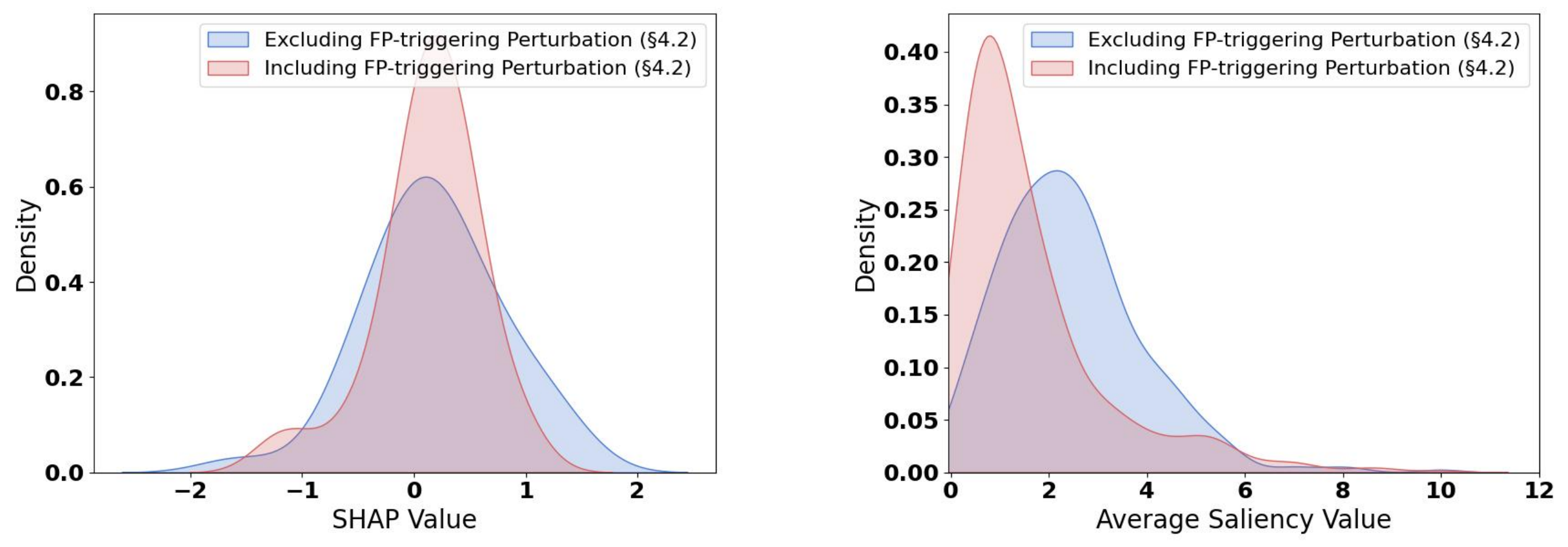}}
    \caption{
    Kernel density estimates of Shapley (left) 
    and saliency (right) values for 1,000 
    randomly selected functions 
    from 10 baseline binaries 
    (100 functions per binary). 
    The blue distribution represents 
    the model’s original behavior 
    (excluding FP-triggering instructions), 
    while the red distribution indicates
    its response after introducing 
    the perturbation.
    On the left, the Shapley values 
    shift upward toward zero, indicating 
    the model increasingly perceives 
    the two functions as similar.
    On the right, the shift toward lower 
    saliency values suggests that 
    the model's attention to 
    crucial features has diminished 
    under the perturbation.
    }
    
    \label{fig:11_xai}
\end{figure}

\PP{FP-Trigger Case Analysis with SHAP}
Shapley values, rooted in 
cooperative game theory, 
quantify the contribution of each token to 
a model’s prediction (\ie similarity 
score by BinShot).
Negative values indicate 
a token’s influence toward 
a dissimilar prediction, while positive values 
indicate influence toward a similar prediction.
\autoref{fig:11_xai} (left) illustrates the 
distribution of Shapley values 
before and after applying 
the FP-triggering perturbation.
Initially, the values span a broad range, 
but after perturbation, most values 
converge near zero, leading to
a false positive.
This concentration around zero 
suggests the model’s decision 
lies close to the decision boundary.
We hypothesize that this is because
our sampling algorithm terminates
once a false positive is detected.
%

\PP{FP-Trigger Case Analysis with a Saliency Value}
As illustrated on the right side in~\autoref{fig:11_xai}, 
we assess instruction-level importance 
using saliency scores: 
computing the gradient of 
the loss with respect to 
each instruction’s embedding and 
taking its norm as the saliency score.
A higher gradient norm indicates 
greater importance, as small changes 
to the embedding would significantly 
affect BinShot’s output. 
Conversely, lower saliency suggests 
that the model places less emphasis 
on those instructions.
As the red curve indicates, 
the introduction of FP-triggering 
instructions causes a noticeable shift 
toward lower saliency values, indicating 
that the model becomes less attentive 
to key features.
This aligns with the model being misled 
into perceiving the two functions 
as more similar than they actually are.

\begin{figure*}[!t]
    \centering
    \resizebox{0.90\textwidth}{!}{
    \includegraphics[width=\textwidth,  trim=30 173 40 7.5, clip ]{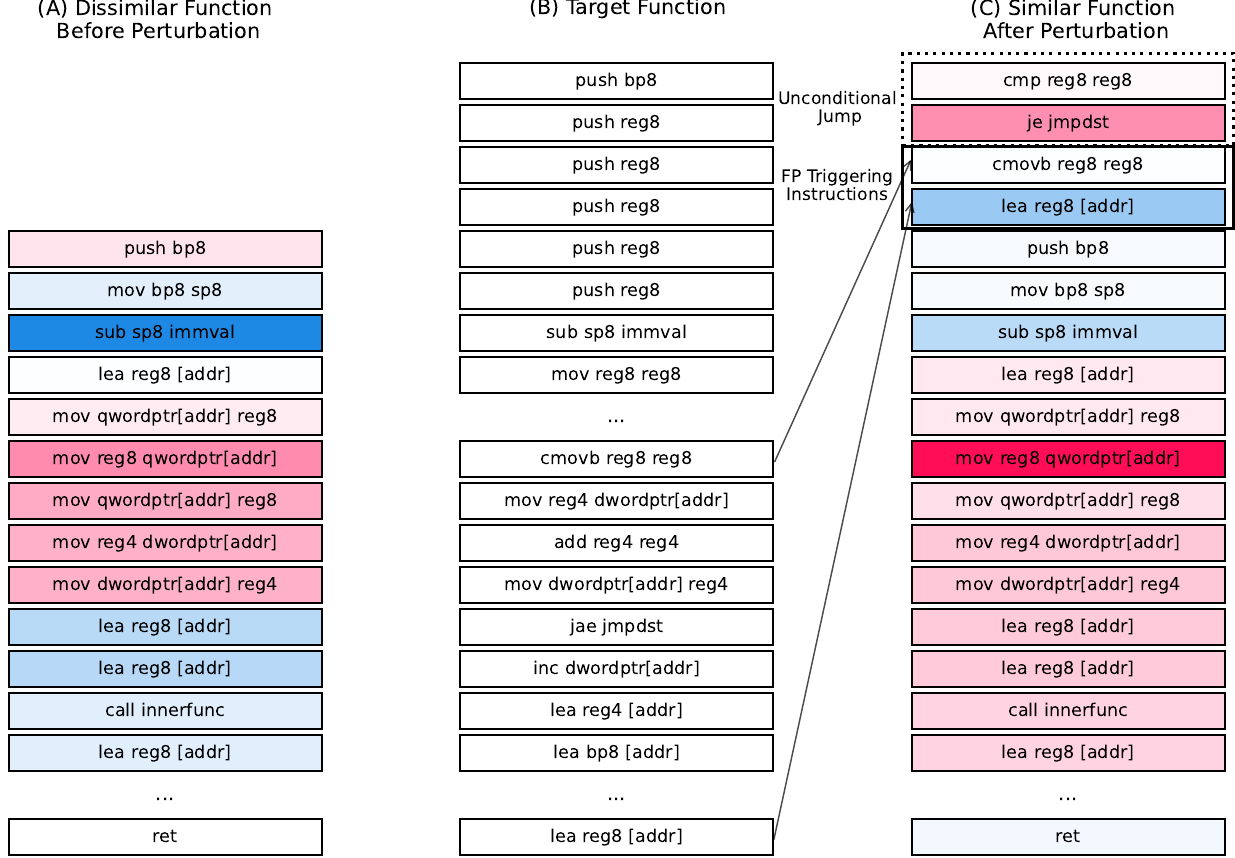}}
    \caption{
    Visualization of instruction-level Shapley 
    values for a victim function 
    (\cc{etarldouble} from \cc{gcc}) 
    and before (\cc{A}) 
    and after (\cc{C}) 
    an FP-triggering perturbation.
    The perturbation modifies the victim function 
    (\cc{A} $\rightarrow$ \cc{C}) to resemble the 
    target function (\cc{vsf\_sysutil\_sendfile} 
    from \cc{vsftpd}, \cc{B}) by inserting 
    FP-triggering instructions sampled 
    using our greedy sampling algorithm
    (Section~\ref{sss:fpt-perturb})
    from the target function (solid box),
    which preserves the original code
    semantics via an unconditional jump (dotted box).
    The red and blue colors highlight instructions 
    contributing to ``similar'' and ``dissimilar'' 
    classifications, respectively.
    The increase in red-labeled instructions 
    after perturbation explains 
    how the false positive has been derived; 
    \eg the four blue instructions 
    in the middle (three \cc{lea reg8 [addr]} 
    and one \cc{call innerfunc} instructions) 
    have been changed to red ones.
    Note that we simplify the 
    functions 
    for brevity.
    } 
    \label{fig:12_xai}
\end{figure*}

\PP{In-depth Analysis on FP-triggering Perturbation}
\autoref{fig:12_xai} illustrates 
the Shapley values
associated with our 
FP-triggering perturbation.
Instructions are color-coded 
by their Shapley values: 
red indicates a contribution 
toward the ``similar'' label, 
while blue represents a contribution 
toward ``dissimilar''.
The figure showcases a 
representative target–victim pair: 
\cc{vsf\_sysutil\_sendfile} 
from \cc{vsftpd} and \cc{etarldouble} 
from \cc{gcc}. 
In this case, the victim function 
is perturbed to resemble 
the target, causing the model 
to return a false positive.
We present the Shapley attributions 
both before and after the perturbation, 
using a normalized instruction 
representation~\cite{koo2023binary}, 
rather than raw x86 assembly,
which follows the pre-processing
in BinShot~\cite{ahn2022practical}.
The increase in red-highlighted instructions 
after perturbation confirms the model’s 
perception has been shifted, 
leading to the false positive.

\subsection{Summary of Findings}
With extensive experiments on evaluating the robustness
of (ML-based) BCSD models against 
various semantics-preserving transformations, 
we discover a few noteworthy findings.
First, the robustness of a model
heavily relies on the pipeline of building the model 
from code pre-processing (\eg normalization, embedding means), 
model architecture (\eg Transformer, BERT), to selecting 
internal features (\eg static, dynamic).
Static features encompass 
the embeddings and the order of an instruction 
(\ie opcode and operands) whereas dynamic features 
include a CFG, a call graph,
and execution traces.
These factors collectively form the ingredients when training a model,
allowing each vector to represent varying code semantic features.
For example, code transformations based on a memory layout
such as ICT and inter-BBR thwart BinShot~\cite{ahn2022practical} that lacks the information.
Likewise, code obfuscations such as control flow flattening and bogus control flow
defeat Gemini~\cite{xu2017neural} and Genius~\cite{feng2016scalable} that depend on the CFG,
which BinShot has shown to be robust against.
This is even clear when looking into the differences in
performance degradation between implanting a semantic NOP 
and inserting a junk code for Gemini and Genius.
Namely, they are robust against the former because it maintains
a CFG; however, less robust against the latter 
because of the alteration of the graph.
Interestingly, in the case of Asm2Vec~\cite{ding2019asm2vec}, its random walk 
exhibits resiliency against graph manipulation to some extent.
It is worth mentioning that quantitatively measuring 
a model's resiliency is difficult; however, we can 
indirectly deduce it throughout its outputs.
Second, an adversary's capability would 
be also bounded by
a transformation budget due to the target model's property
in terms of input length and the expressivity of 
semantically equivalent machine instructions.
Consequently, our work alludes 
to considerations 
for designing
a reliable and robust BCSD model
(but applied to any ML-based model),
which takes machine instructions as input.
%
%
Third, unlike semantics-preserving 
transformations, FP-triggering perturbations 
demand a carefully designed 
sampling strategy, similar to 
adversarial attacks in deep learning.
We discover that FP-triggering 
perturbations are highly effective, 
even with minimal changes, and 
exhibit strong transferability, particularly 
to models with similar architectural characteristics.
Our further investigations
using XAI methods reveal that these 
perturbations disrupt the model’s ability 
to identify essential tokens,
thereby leading to false positive predictions.

\section{Related Work}
%

\PP{Attacks against Deep Learning Models}
Szegedy~\etal~\cite{szegedy2013intriguing} are 
the first to demonstrate that 
deep learning
models are vulnerable 
to adversarial attacks.
Following the exploration, the fast gradient 
sign method~\cite{goodfellow2014explaining} 
introduces a technique for generating 
adversarial noise by perturbing inputs 
in the direction of the gradient 
to maximize prediction error, raising awareness 
about the susceptibility of 
deep learning
models to such inputs.
The projected gradient descent 
attack~\cite{madry2017towards} extends 
the fast gradient sign method by 
iteratively applying gradient updates 
and projecting the perturbation back 
into a constrained region 
to maintain it within a specified norm bound.
Carlini~\etal~\cite{carlini2017towards} 
further expand the attack landscape by 
introducing optimization-based attacks under 
different distance metrics 
(\ie $L_0$, $L_2$, $L_\infty$), enabling 
fine-grained control over perturbation similarity.
Although adversarial attacks initially 
focus on image classifiers, the rise of 
transformer-based models has led to 
similar vulnerabilities in NLP.
TextFooler~\cite{jin2020bert} and 
BERT-attack~\cite{li2020bert}, for instance, 
generate adversarial examples by replacing 
words with synonyms, often causing models 
to misclassify inputs such as sentiment 
or intent, which inspires our work.
Meanwhile, Zou~\etal~\cite{zou2023universal} 
introduce a universal adversarial suffix that 
can manipulate LLMs into generating harmful 
content when appended to prompts.

\PP{Perturbation Attacks against Executable Binaries}
Lucas~\etal~\cite{lucas2021malware} propose
an adversarial attack that perturbs 
malicious PE executables via
binary instrumentation.
Their approach combines in-place code 
transformation with a modified displacement 
that integrates semantic NOPs to evade detection.
Similarly, our experiments incorporate 
both in-place code transformation and 
semantic NOP implantation as part 
of our adversarial variant generation strategy.
%

\PP{Attacks against BCSD Models}
FuncFooler~\cite{jia2022funcfooler} 
demonstrates a practical black-box attack 
on learning-based BCSD models, including 
SAFE~\cite{massarelli2019safe}, 
Asm2Vec~\cite{ding2019asm2vec}, 
and jTrans~\cite{wang2022jtrans}. 
Capozzi~\etal~\cite{capozzi2023adversarial} 
extend this line of work
by distinguishing between targeted attacks
(\ie designed to 
make dissimilar functions appear similar) 
and untargeted attacks
(\ie aiming to make similar functions 
appear dissimilar) across
three models: Gemini~\cite{xu2017neural}, 
Genius~\cite{feng2016scalable}, 
and SAFE~\cite{massarelli2019safe}.
While both FuncFooler and 
Capozzi~\etal\footnote{Unfortunately, 
a direct comparison was not possible, as
neither approach was publicly available 
at the time of writing.}
share the goal of evaluating BCSD model 
robustness, their approaches are limited 
to a single semantics-preserving transformation 
implemented via inline assembly 
for source code perturbation.
In contrast, our work explores eight 
distinct semantics-preserving transformations, 
most applied at the binary level, without 
modifying any source code.

\section{Discussion and Limitations}

\PP{Formal Verification}
The in-place code transformation 
techniques proposed by Pappas~\etal~\cite{pappas2012smashing} 
are not formally verified.
As a result, we do not guarantee 
full correctness across all code instances, 
although we partially validate that 
each transformation behaves as intended.

\PP{Variant Exploration}
Moreover, both inter-basic-block reordering 
and in-place code transformation are 
non-deterministic, which may impact 
the observed robustness of each model.
While exhaustively generating all possible variants 
is infeasible, our study still yields meaningful 
insights into model behavior.
We fix the insertion point of semantic 
NOPs and junk code at the function entry 
to isolate the impact of sequence length.
Although such sequences could be placed elsewhere,
placing them at the beginning exploits the positional bias
of language models, which prioritize information 
appearing early in long contexts~\cite{lostmiddle}.
%
%
%

%

\PP{XAI Fidelity}
Lastly, while we analyze FP-triggering transformations 
using SHAP~\cite{lundberg2017unified} and 
saliency~\cite{saliency}, prior work has shown 
that XAI techniques can yield 
conflicting results~\cite{he2023good}.
A comprehensive fidelity analysis 
of XAI methods is beyond the scope of this study.


\PP{Other Binary Code Obfuscations}
Beyond the code obfuscation techniques 
in our experiments, many other methods exist,
such as inserting trampolines (\ie jump tables) 
or applying string and data obfuscation.
Additionally, our evaluation does not consider 
advanced protections like encryption, 
code packing, and compression 
(\eg self-extracting or polymorphic code), 
nor commercial obfuscation tools 
such as Themida~\cite{thermida} 
or VMProtect~\cite{vmprotect}.
We leave the exploration of 
additional obfuscation techniques 
as part of future work.

\PP{Defenses against Adversarial Code Transformations}
Adversarial samples are 
deliberately crafted to mislead an ML model and induce errors.
Data augmentation
with known transformation suites
can be helpful to increase the robustness of 
an ML-based BCSD model so that the model can learn varying code
representations from variants.
Another direction is training that combines static features
(\eg CFGs, call graphs, API-call graphs, data flows) with dynamic
features (\eg API traces, syscall traces, behavior traces) can assist in
the model's final decision.
We leave a systematic study of adversarial training 
with semantics-preserving transformations to
our future work.


\section{Conclusion}
%
%
Recent advancements in deep learning offer promising 
opportunities to enhance binary analysis for security applications.
Despite the surge of various models available,
their resiliency against adversarial samples has not been studied in depth.
In this paper, we focus on evaluating the robustness 
of six state-of-the-art
binary code similarity detection
models with
eight semantics-preserving code transformations.
Our major finding highlights that model robustness 
significantly depends on the particular characteristics of the processing pipeline, 
including code preprocessing, model architecture, 
and feature selection.
Furthermore, the effectiveness of semantics-preserving
transformations is limited both by
inherent constraints of the model and by the expressive capacity
of semantically equivalent instructions within the binary.

\section{Data Availability}
We have open-sourced \sys\footnote{https://zenodo.org/records/17116512} 
and our adversarial binary dataset 
for further exploration of model robustness 
in machine-learning-based binary analysis.

\section*{Acknowledgments}
We thank the anonymous reviewers 
for their constructive feedback.
This work was partially 
supported by grants from the
Institute of Information \& Communications 
Technology Planning \& Evaluation (IITP),
funded by the Korean government 
(MSIT; Ministry of Science and ICT): 
No. RS-2024-00337414, 
and
No. RS-2024-00437306. 
Additional support was provided by 
the Basic Science Research Program through
the National Research Foundation of Korea (NRF),
funded by the Ministry of Education
of the Government of South Korea: No. RS-2025-02293072.
Any opinions, findings, and conclusions or 
recommendations expressed in
this material are those of the authors and 
do not necessarily reflect
the views of the sponsor.


\footnotesize
\bibliographystyle{ACM-Reference-Format}
\bibliography{references}

\end{document}